\documentclass{article}
\usepackage{FPSpro}
\usepackage{epsfig}
\def\etal{{\sl et al.}}

\def\rhobar{\bar{\rho}}
\def\etabar{\bar{\eta}}
\def\bag{\hat B}
\begin{document}
\title{ 
  CKM Status and Prospects
  }
\author{
  Brian K. Heltsley  \\
  {\em Laboratory of Nuclear Studies, Cornell University, Ithaca, NY 14853} \\
\\
{\em To appear in the proceedings of the}\\
{\em XXI Physics in Collision Conference}\\
{\em Seoul, Korea, 28-30 June, 2001}
  }
\maketitle
\baselineskip=11.6pt
\begin{abstract}

  This presentation reviews the Standard Model formalism governing 
the weak decays of quarks, as embodied in the Cabibbo-Kobayashi-Maskawa 
(CKM) weak mixing matrix, and summarizes the experimental status and outlook.
Two recent CLEO $|V_{cb}|$ analyses are described. 
Complications from strong interactions in relating the experimentally 
accessible quantities to CKM elements are highlighted, and prospects
for addressing those difficulties with LQCD and CLEO-c are outlined. 

\end{abstract}
\baselineskip=14pt
\section{Introduction}

  We gather at an auspicious time for particle physics in general, and
heavy flavor physics in particular. The highly successful LEP program 
has concluded, making way for the LHC. Run II at the Tevatron is 
underway. Efforts to complete the $B$-factories and harvest their
results have nearly come to fruition: KEK-B and PEP-II have 
achieved luminosity exceeding expectations, and Belle and BaBar 
both presented initial results last summer. It is widely anticipated 
that, in the coming weeks, direct and definitive 
measurements of non-zero CP violation in the $B-$meson system will 
be announced by one or both $B-$factory collaborations. Such a 	
development would undoubtedly intensify the spotlight on the heavy 
quark sector of the Standard Model. 

  Just two days prior to the opening of this conference, the CESR/CLEO 
$B-$physics program concluded two decades of data-taking, marking a 
shift of their focus from $B$'s to $D$'s, $\psi$'s, and $\Upsilon$'s. 
This shift is embodied in the imminent CESR-c/CLEO-c project, which will
have unique capabilities to pursue precision charm measurements and 
QCD tests that are essential to our understanding of the the heavy 
flavor sector.

  Activity and broad interest in the weak decays of heavy quarks 
and related issues has never been higher. The landscape is changing
rapidly. This review provides a snapshot of that landscape 
as of early summer 2001.

\section{CKM, the Unitarity Triangle, and $CP$ Violation}
  In the Standard Model, the charged current weak interactions of three 
generations of quarks are governed by a Lagrangian which contains a 
transformation from the mass eigenbasis to the flavor (generation) 
eigenbasis.\cite{bib:PDG,bib:BABPHYS,bib:BURAS2} 
This flavor-mixing is expressed 
as a 3$\times$3 complex matrix $V$ known as the
Cabibbo-Kobayashi-Maskawa (CKM) matrix.\cite{bib:CAB,bib:KM} 
The elements $V_{ij}$ determine the relative weak couplings of the charge 
$-{2\over3}$ quarks $i=(u,c,t)$, one row each, to the 
charge $+{1\over3}$ quarks ($j=d,s,b$), one column each. By
definition, the matrix $V$ is {\bf unitary}. 
Unitarity reduces the number of independent parameters to nine, which 
can be chosen as three real mixing angles and six imaginary phases. Five of
the phases are removable. The four remaining parameters are
fundamental constants of nature to be determined by experiment; the
Standard Model itself gives no guidance as to their values.

  The successive application of charge conjugation ($C$), parity
reversal ($P$), and time reversal ($T$) is an exact symmetry of
any local Lagrangian field theory. The electromagnetic and strong
interactions preserve all three symmetries separately; weak interactions
violate $C$ and $P$ separately but together appear to preserve $CP$ with the 
exception of neutral kaon (and perhaps beauty) decays. Within
the Standard Model framework, $CP$ violation can occur only if the one 
irremovable CKM phase is non-zero. $CP$ violation would not happen, 
however, if any two same-charge quarks have equal masses, if any mixing 
angle is $0$ or $\pi\over2$, or if the sine of the phase is zero.

  It is instructive to note a couple of special cases: first, 
if applied to the leptonic sector, in the limit of vanishing neutrino 
masses, the mixing matrix is trivially the identity (i.e. no
mixing and no $CP$-violation); second, the unitarity constraint
in the case of just two quark generations leaves no irremovable phase
and hence no $CP$ violation. The latter property of $V$ and the 
existence of $CP$ violation in neutral kaon decay provoked the 
hypothesis of a third generation\cite{bib:KM} before its experimental 
discovery. With such a history, it is not surprising that experimental 
unitarity tests continue to be pursued vigorously.

  The unitarity constraint imposes unity normalization on all rows
and columns as well as their orthogonality. The orthogonality 
relations can be described geometrically by six 
``unitarity triangles''
in the complex plane: each side of the $jk$ triangle is
a vector $V_{ij}V^*_{ik}$. All six triangles have the
same area, which is non-zero only if $CP$ violation occurs. 

\begin{figure}
  \vspace{7.5cm}
  \includegraphics{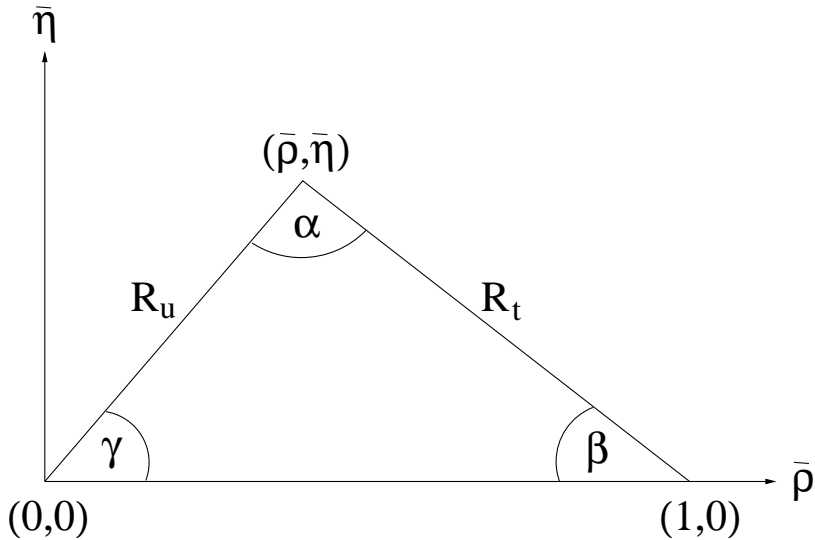}
  \caption{\it
    The rescaled Unitarity Triangle in the Wolfenstein-Buras
  formulation, in which $R_x = |V_{xd}V_{xb}^*/(V_{cd}V_{cb}^*)|$.
   [from ref.~8]
    \label{fig:ut} }
\end{figure}

\begin{figure}[htb]
  \vspace{10.2cm}
  \includegraphics{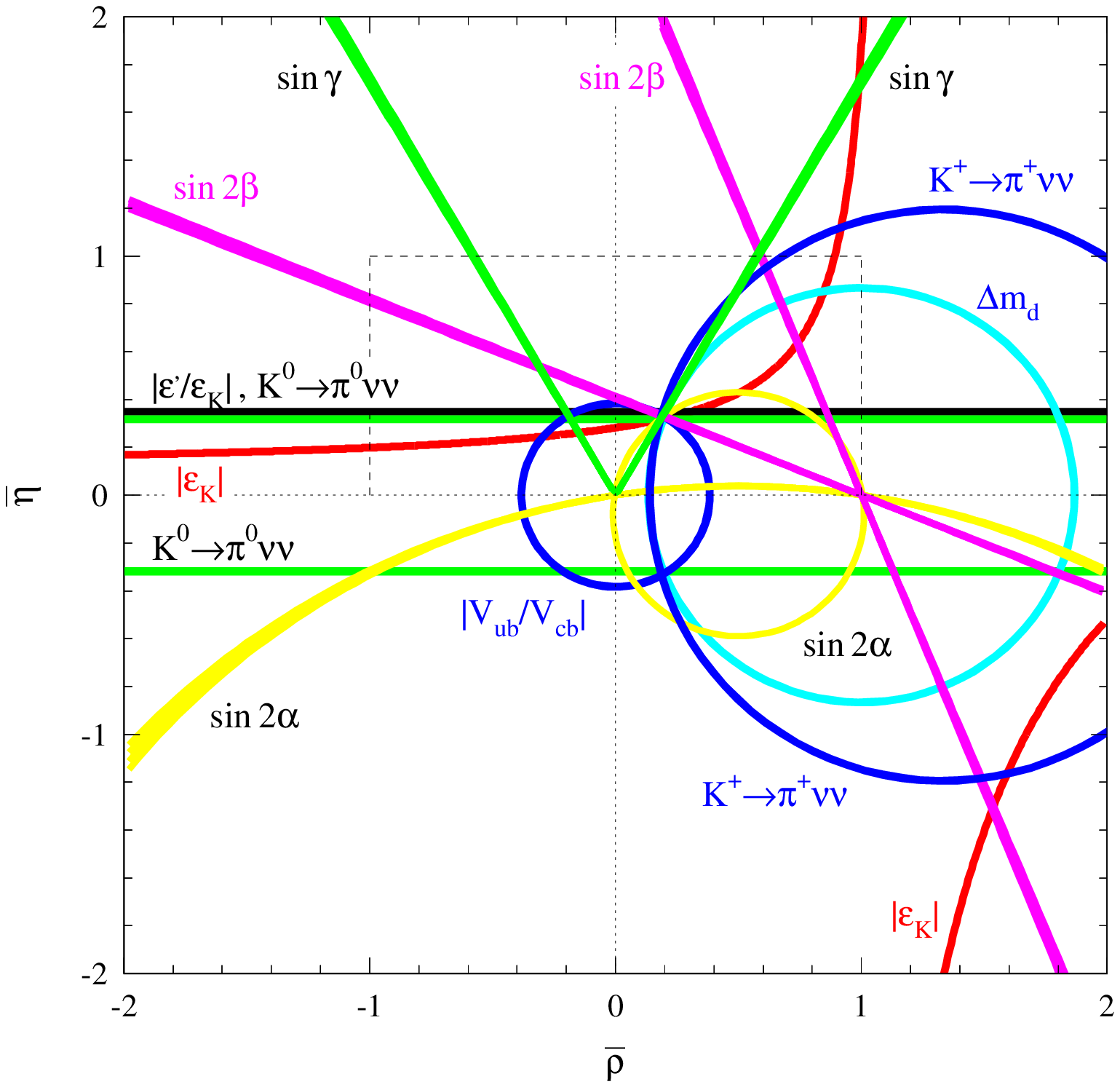}
  \caption{\it
     Contours in the $\rhobar,\etabar$ plane corresponding to 
     hypothetical precision measurements of nine observables,
     arranged to be consistent with unitarity by overlapping at
     a fixed (arbitrary) point $\sim (0.2,0.35)$.
   [from ref.~8]
    \label{fig:ckmhypo} }
\end{figure}

  With the $\sim$1\% knowledge from experiment of 
$|V_{cd}|\approx |V_{us}|=\lambda\equiv\sin\theta_C=0.22$,
it can be shown that four of the unitarity triangles ($ds$, $sb$, $uc$,
$ct$) have one side that is smaller by two or more powers of
$\lambda$ than the other two sides. The $db$ and $ut$ triangles have 
three comparable sides of length $\sim\lambda^3$. The $db$ triangle,
also known as {\sl the} Unitarity Triangle (UT), is described by
\begin{equation}
V_{ud}V_{ub}^* + V_{cd}V_{cb}^* + V_{td}V_{tb}^* = 0.
\label{eq:UT}
\end{equation}
By convention, $V_{cb}^{*}V_{cd}$ is used to normalize the other 
UT sides, forming the {\sl rescaled} UT shown in Fig.~\ref{fig:ut}. 
Of the six triangles, the one described by Eq.~\ref{eq:UT} has
attracted the most attention due to the relative accessibility of
sides and angles to experiment. Several measurements can be used in 
combination to directly test unitarity, and hence the validity of 
the Standard Model itself.

  One favored explicitly unitary parameterization of $V$ has 
three mixing angles and a phase.\cite{bib:PDG}
Generational symmetry and therefore ease of interpretation are preserved
in this representation. However, currently prevalent
is the Wolfenstein\cite{bib:WOLF} formulation
as modified by Buras\cite{bib:BURAS1}
\begin{equation}
V
   =\left(\begin{array}{ccc}
1 - {{\lambda^2}\over 2} & \lambda & A\lambda^3(\rho - i\eta ) \\
-\lambda-iA^2\lambda^5\eta & 1 - {{\lambda^2}\over 2} & A\lambda^2 \\
A\lambda^3(1-\rhobar-i\etabar ) & -A\lambda^2 - iA\lambda^4\eta & 1 
          \end{array}\right)
\label{eq:CKMWB}
\end{equation}
\noindent
in which $\lambda\equiv {\rm sin}\theta_{C}\approx 0.22$ has become an 
expansion parameter, and terms of order $\lambda^6$ ($\sim 10^{-4}$)
and higher are dropped. The Buras corrections, addition of the terms of 
higher order than $\lambda^3$ and the rescaling of $\rho$ and $\eta$ in 
$V_{td}$ to $\rhobar\equiv\rho (1-{{\lambda^2}\over 2})$ and 
$\etabar\equiv\eta(1-{{\lambda^2}\over 2})$, attain better precision 
and exact unitarity. The four fundamental couplings are then $\lambda$, 
$A$, $\rhobar$, and $\etabar$. Here the rescaled UT has vertices in the 
complex plane at $(0,0)$, $(1,0)$, and $(\rhobar,\etabar)$ as in
Fig.~\ref{fig:ut}. The attraction of 
this representation is that the hierarchy of matrix element sizes in 
powers of $\lambda$ is explicit, as is the presence of the UT coordinates 
$(\rhobar,\etabar)$ as separate fundamental constants. The internal 
angles $\beta$, $\gamma$, and $\alpha$ (or $\phi_1$, $\phi_3$, $\phi_2$,
respectively) can be taken as an alternative basis for CKM 
parameters.\cite{bib:SILVA}

 $CP$ violation occurs only if $\etabar$ (and therefore $\beta$) are 
non-zero. Within the Standard Model framework, measurement of 
$CP$-conserving decays can determine the lengths of the UT sides, 
and therefore {\sl indirectly} the value of a non-zero $CP$-violating 
$\beta$. However, the three individual angles and their sum must all
be subjected to measurement for consistency with unitarity to test
the limitations of the Standard Model itself. Eventually, all the 
unitarity triangles need to be probed for consistency among sides
and angles; some will be easier and/or cleaner than others. 

Global fitting can improve the metrological
accuracy of all the CKM couplings. Fig.~\ref{fig:ckmhypo} shows the
contours that would result from hypothetical precision measurements
of nine observables, the values of which were chosen to be
consistent with unitarity and therefore all intersect at a point.
The reality now is that only a few of those quantities have
been determined at all, and, for the most part, not very precisely.

\section{Complications from Strong Interactions}
\label{sec:QCD}

  Because hadrons, not quarks, are produced and observed in the 
laboratory, CKM matrix elements generally cannot be extracted 
from experiment without understanding the effects of strong 
interactions on the initial and/or final state particles. The 
required computational alacrity must come from quantum chromodynamics 
(QCD), the gauge theory of quarks and gluons. Although modeled after
QED, QCD differs in the non-linearity of its field equations (because 
gluons carry color charge), and the permanent confinement of quarks inside
hadrons (due to strong coupling). Asymptotic freedom has allowed
the development of perturbative QCD, which has been very
successful in describing high momentum-transfer (QED-like) processes
such as deep inelastic scattering (DIS) and high energy quark and gluon jets.
However, untangling CKM matrix elements from the weak interactions of 
hadrons frequently requires theoretical understanding of the low energy, 
non-perturbative regime. Such understanding can be described presently
as incomplete at best.

  Lattice QCD (LQCD), Heavy Quark Effective Theory (HQET), the
Operator Product Expansion (OPE), and several models all currently 
enjoy varying degrees of success in calculating effects of strong
interactions. HQET predictions are 
rigorously accurate in the limit of infinitely heavy quark masses, 
and hence must be corrected in powers of $1/M_q$.
Models are generally tuned to reproduce one set of observations
and then applied to predict others. LQCD, however, is a complete definition
of the theory. Although unproductive for most of its existence,
LQCD has recently overcome technical problems\cite{bib:LEPAGE} and 
adopted new algorithms\cite{bib:ALFORD} which, along with availability 
of inexpensive computing power, offer the hope of $\sim$1\% precision 
on dozens of quantities within the next five years. Already some form 
factors, decay constants, and glueball masses have LQCD predictions 
estimated to be accurate at the 10-15\% level. 

  Quantities needed from non-perturbative QCD include decay constants,
form-factors, and bag parameters. The decay constant appears 
in the leptonic decay of a meson $Q$ as
${\cal B}(Q\to l\nu)\sim m_l^2 f_Q^2 |V_{qj}|^2 \tau_Q$. Unfortunately
such branching ratios of $D$'s and $B$'s are small enough to make
purely leptonic decays not useful at present.   Form-factors 
$F_X(q^2)$ play an analogous role in the $q^2$-dependence of 
semi-leptonic decays 
$d\Gamma/dq^2(Q\to Xl\nu)\sim p_X^3 |V_{qj}|^2 |F_X(q^2)|^2$ and
are specific to the hadronic final state $X$. The $q^2$ dependence
is helpful independently from the overall normalization 
because the form-factor value at zero recoil ($q^2$=0) is accessible 
to HQET; one can extrapolate the measured $F(q^2)|V_{qj}|$
to $q^2=0$ to extract $|V_{qj}|$. The bag parameter appears 
with the decay constant squared in $f_Q^2\bag_Q|V_{tj}|^2$, the 
expression for the frequency (neutral
meson mass difference) of 
$Q\iff\bar{Q}$ oscillations  (i.e. $K^0$, $B_d$ and $B_s$ mixing), 
which are dominated for $B$'s by box diagrams with 
$t$ and $W$ sides.

   Inclusive measurements of semi-leptonic decays of charm and bottom 
are employed to avoid the need for form-factors, relying on HQET for 
the necessary quark-level input. These results rest upon the assumption 
of {\sl quark-hadron duality}, which presumes that integrated over a 
broad enough spectrum of hadronic final states and phase space, the measured 
quantity is the same as the quark-level prediction. Some take this 
duality as a given, especially for unrestrictive event selections; some 
express reservations and wonder how to assess
uncertainties when experimental cuts are quite narrow, and some 
take inclusive results as measurements not of CKM parameters but 
of the validity of the duality assumption itself. In any case,
inclusive measurements can be instructive as to the underlying physics
and assumptions.

\section{CKM Status without Unitarity}
 What do we know about the CKM elements now, without unitarity
constraints? Of the nine absolute values, two are measured to $\le$1\%, 
three more are known to $\sim$6\%, and the remaining four are 
uncertain at the $\ge$15\% level. Measurements of super-allowed and neutron 
$\beta$-decays\cite{bib:PDG,bib:TOWNER} yield 
$|V_{ud}|\approx 0.974\pm 0.001$. Analysis\cite{bib:PDG,bib:LEUT} 
of neutral and charged $K_{e3}$ decays ($K\to \pi e\nu$) yields 
$|V_{us}|\approx 0.2196\pm 0.0023$, which is verified by a somewhat 
less precise result from hyperon semi-leptonic 
decays.\cite{bib:FLORES} Di-muon production in DIS of $\nu$ and $\bar{\nu}$ on nucleons, one muon from 
a semi-leptonic charm decay, has been 
examined\cite{bib:CDHS}, yielding 
$|V_{cd}|=0.224\pm0.016$ and $|V_{cs}|=1.04\pm0.16$. From direct 
production of charm in real $W$ decays,\cite{bib:OPAL} 
$|V_{cs}|=0.969\pm 0.058$, considerably more accurate than DIS.
$|V_{cs}|$ from $D\to Ke\nu$ decays\cite{bib:PDG} is consistent 
but has an error nearly three times larger, due mostly to uncertainty 
in the form-factor. Measurements of the $b$-fraction in top quark 
decays by CDF and D0 result in the rather loose restriction\cite{bib:PDG} 
of $|V_{tb}|<(0.99\pm 0.15)\times U_t$ where $U_t^2=\sum_j{|V_{tj}|^2}$.

  The remaining CKM elements (third row and column) can be measured
with $B$-decays. 
Currently there is information 
from both inclusive and exclusive semi-leptonic decay branching ratios 
for  $V_{ub}$ and $V_{cb}$. 
The heavy-to-heavy transition in inclusive $b\to cl\nu$
is more easily calculable in HQET than for $b\to ul\nu$; the published
numbers are $|V_{cb}|=(40.7\pm0.5\pm2.0)\times 10^{-3}$ from 
LEP\cite{bib:LWGVCB} and 
$|V_{cb}|=(41\pm2\pm2)\times 10^{-3}$ from CLEO,\cite{bib:CLEOVCB} in which the
errors are experimental and theoretical, respectively. Both are subject to the
inherent assumption of quark-hadron duality. Exclusive results
from LEP and CLEO for $|V_{cb}|$ also give results in this range, 
as seen in Section~\ref{sec:bcex}.

  $|V_{ub}|$ is difficult due to the enormous charm
background over all but the lepton-momentum spectrum's endpoint region, 
and the difficulty of obtaining the requisite QCD information to extract
it from the branching fraction. Two different strategies 
have emerged. The inclusive LEP analysis includes a wide kinematic 
range to avoid losing signal statistics but then pays the price of 
quark-hadron duality and a fine-tuned modeling of charm backgrounds.
CLEO restricts itself to exclusive final states ($B\to\pi l\nu$, $\rho l\nu$) 
using $\nu$-reconstruction, in which there is a more 
favorable signal-to-noise but a considerable uncertainty in the 
form-factors. The LEP value\cite{bib:LWGVUB}
$|V_{ub}|=(4.09^{+0.59}_{-0.69})\times 10^{-3}$, in which experimental 
and modeling errors make comparable contributions, is consistent with 
that from CLEO,\cite{bib:CLEOVUB} $(3.25\pm0.30\pm0.55)\times 10^{-3}$, 
in which the first error is experimental and the second is theoretical.

  Measurement of the frequency of oscillations in the $B_d$ and $B_s$
systems yields\cite{bib:LWGBO} 
$\Delta m_d=0.489\pm 0.008$~ps$^{-1}$ and $\Delta m_s>14.6$~ps$^{-1}$. Relating
these to the CKM matrix elements requires knowing $\bag_x f_{B_x}^2$
for each meson. In the case of $B_d$ it can be computed using LQCD
to within $\sim$20\%, yielding $|V_{tb}^*V_{td}|=(8.3\pm 1.6)\times10^{-3}$.
The ratio $\Delta m_s/\Delta m_d$ is more tractable theoretically
than $\Delta m_s$ alone due to the cancelation of several factors
as well as the feature that only the ratio of bag parameters comes into play,
not their absolute values. LQCD has given $\sim$10\% errors in
the ratio $\xi^2=\bag_d f_{B_d}^2/(\bag_s f_{B_s}^2)$, translating the
$\Delta m_s$ limitation to $|V_{td}|< 0.24|V_{ts}|$.

  Two measurements of $CP$-violation put limits on CKM matrix elements.
The relationship of the $CP$-violating parameter $\epsilon_K$ in
neutral kaon decay to CKM elements is limited not by experiment
but by uncertainty in the bag constant $B_K$ which currently has
a $\sim$15\% uncertainty. In contrast, the UT angle 
$\beta\equiv arg(-V_{cd}V_{cb}^*/(V_{td}V_{tb}^*))$  is directly 
sensitive to the magnitude of the asymmetry between mixed and unmixed 
neutral $B_d$ decays to the same color-suppressed $CP$ eigenstate,
suffering only very small theoretical uncertainties. This explains why 
it is the first and foremost target of the 
$B$-factories.\cite{bib:ROODMAN} Although
the world average of sin$2\beta=0.48\pm0.16$ is three standard
deviations from zero, it may be prudent to exercise some caution
in regarding this as observation of $CP$ violation, because no single
experiment as yet has a two-standard deviation effect. However,
with almost no theoretical uncertainty and systematic errors
at the $\pm$0.05 level or better, both Belle and BaBar have the
data in hand for such unambiguous discovery.

\section{New CLEO Results on $|V_{cb}|$ from $D^*l\nu$ and Moments}
\label{sec:bcex}

\begin{figure}[htb]
  \vspace{8.2cm}
    \includegraphics{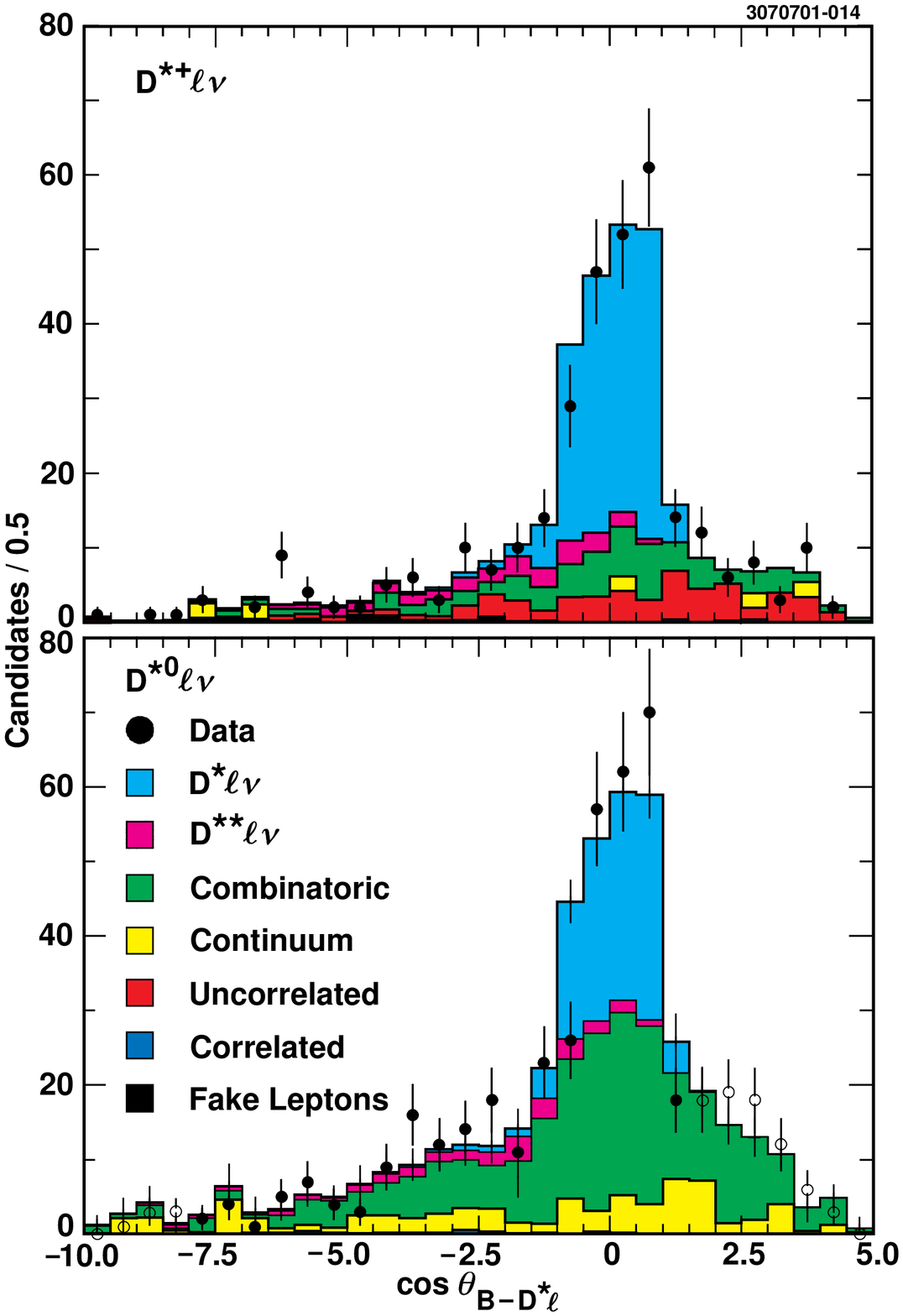}
   
    \includegraphics{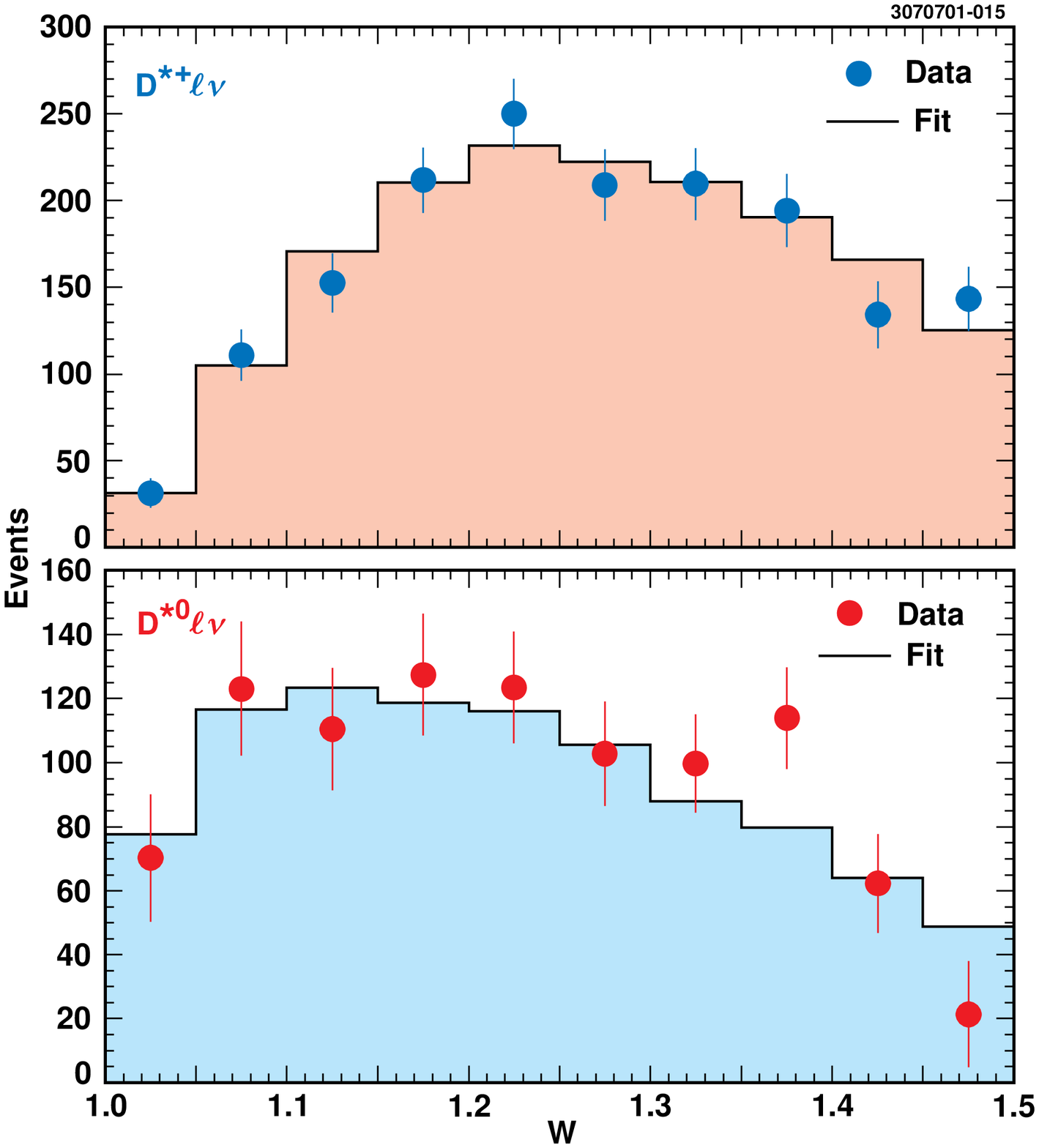}
    \caption{\it
      Distributions from the preliminary CLEO analysis of
      $B\to D^* l\nu$, in the kinematic variable (left)
      used to distinguish among and fit signal and various 
      background contributions in a particular $w$-bin $(1.10-1.15)$; 
      on the right the event yields are shown as a function of 
      $w$ along with the fit of Fig.~\ref{fig:vcb2}.
      \label{fig:vcb1}
      }
\end{figure}

\begin{figure}[phtb]
  \vspace{6.5cm}
    \includegraphics{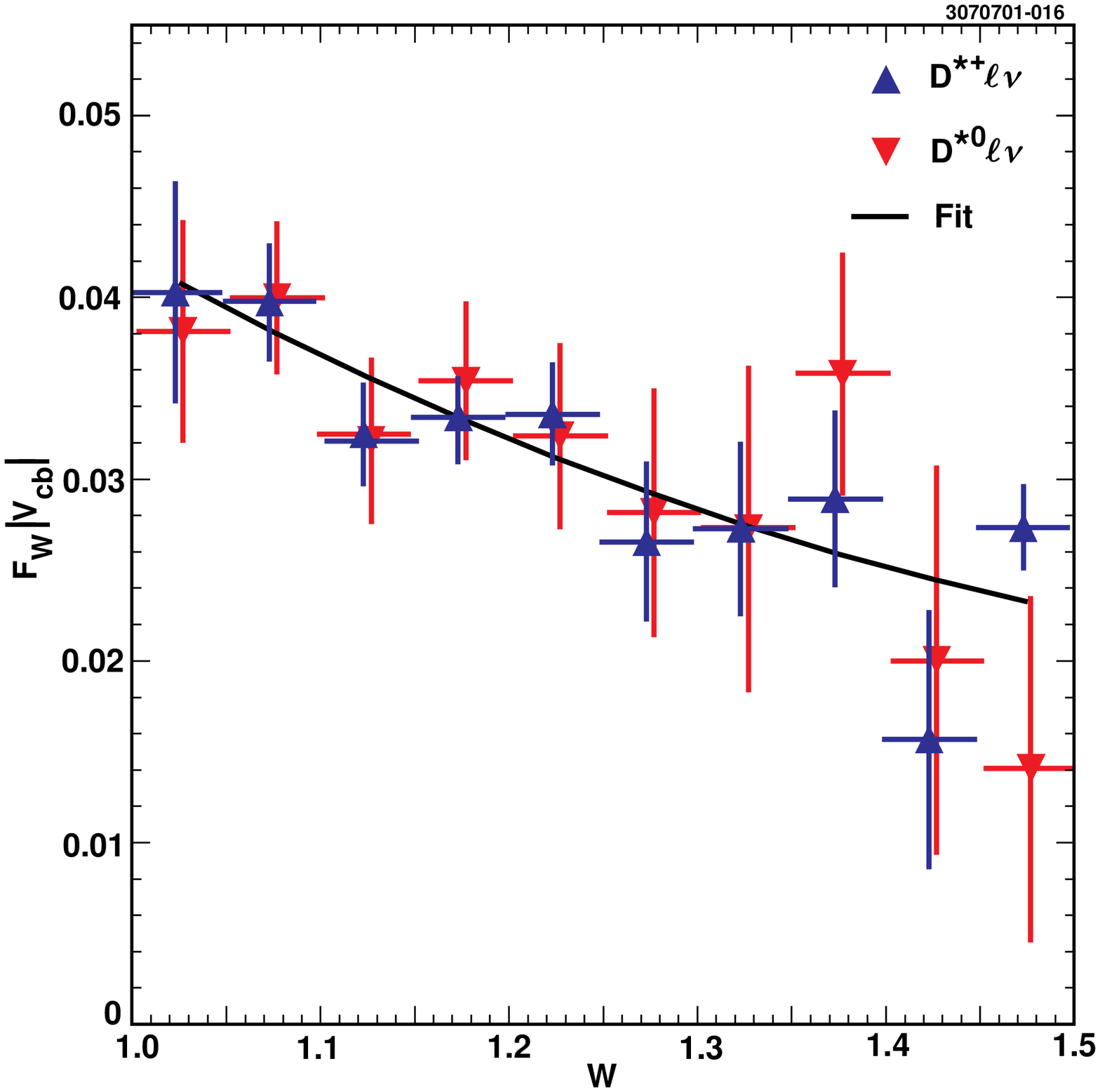}
   
    \includegraphics{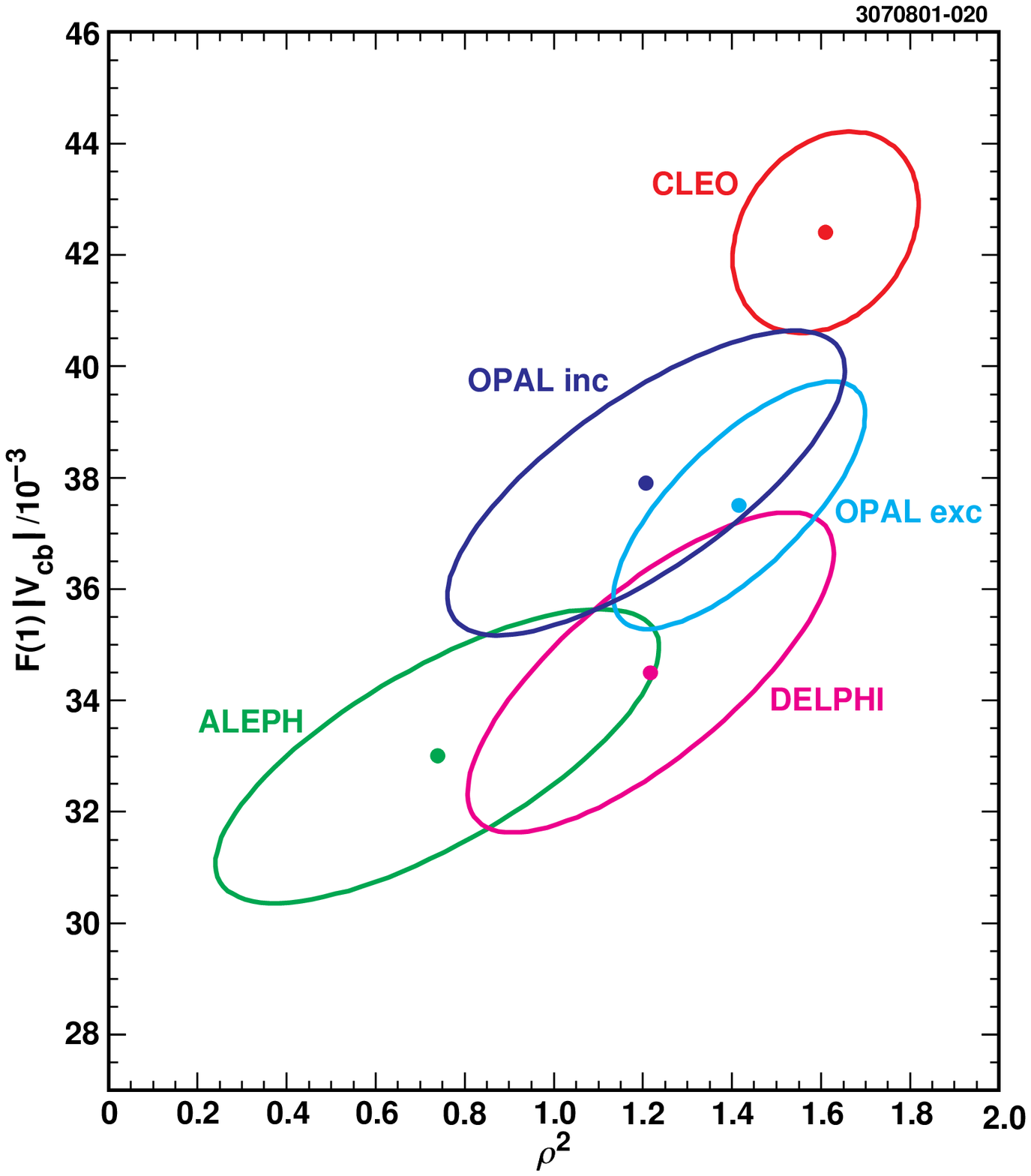}
    \caption{\it
      Left: $F(w)|V_{cb}|$ from the 
      preliminary CLEO $B\to D^* l\nu$
      analysis and resulting fit. Right: $\Delta\chi^2=1$ contours
      in the $\rho^2$-$F(1)|V_{cb}|$ plane for the OPAL, DELPHI,
      ALEPH, and preliminary CLEO $b\to c$ analyses.
      \label{fig:vcb2}
      }
\end{figure}

  CLEO has a new preliminary analysis of $B\to D^*l\nu$ using both
$D^{*+}$ and $D^{*0}$. Both signal and $D^*Xl\nu$ backgrounds are 
determined by fitting data distributions in a kinematic variable sensitive
to their relative contributions. These fits are performed in bins of
a $q^2$-substitute HQET variable $w$, which is the $D^*$ 
boost in the $B$ rest frame and has range $1<w<1.5$. The fits in a 
sample $w$-bin and the overall resulting raw $w$-distributions are 
shown in Fig.~\ref{fig:vcb1}. The efficiency-corrected $F(w)|V_{cb}|$ 
distribution appears in Fig.~\ref{fig:vcb2}. Both CLEO and LEP
extrapolate to the HQET-friendly, zero-recoil value $w$=1 with
the same functional form. The $w$-curvature parameter is $\rho^2$ 
(not to be confused with the CKM parameter $\rho$). The CLEO 
[LEP]\cite{bib:LWGVCB} results
are $F(1)|V_{cb}|=(42.2\pm 1.3\pm 1.8)$ [$(35.6\pm 1.7)$]$\times
10^{-3}$ and $\rho^2=1.61\pm0.09$ [$1.38\pm 0.27$]. CLEO finds
$both$ a smaller $D^*Xl\nu$ background than LEP, which instead of
fitting the data for background 
uses a fixed value from a tuned model,\cite{bib:LEIBO} 
$and$ a larger curvature $\rho^2$. 
These variables are correlated in the final result, as shown in
Fig.~\ref{fig:vcb2}. The LEP and CLEO results are
marginally consistent. Care should be taken in
directly comparing $|V_{cb}|$'s quoted from these analyses, 
as the LEP WG uses\cite{bib:LWGVCB,bib:BSU}
$F(1)=0.88\pm 0.05$ and CLEO prefers $0.913\pm0.042$. 
The CLEO value is more central to the range of predictions, which have
a $\sim$5\% spread. Using these $F(1)$'s makes the $|V_{cb}|$'s seem 
closer than they actually are.

\begin{figure}[phtb]
  \vspace{6.6cm}
    \includegraphics{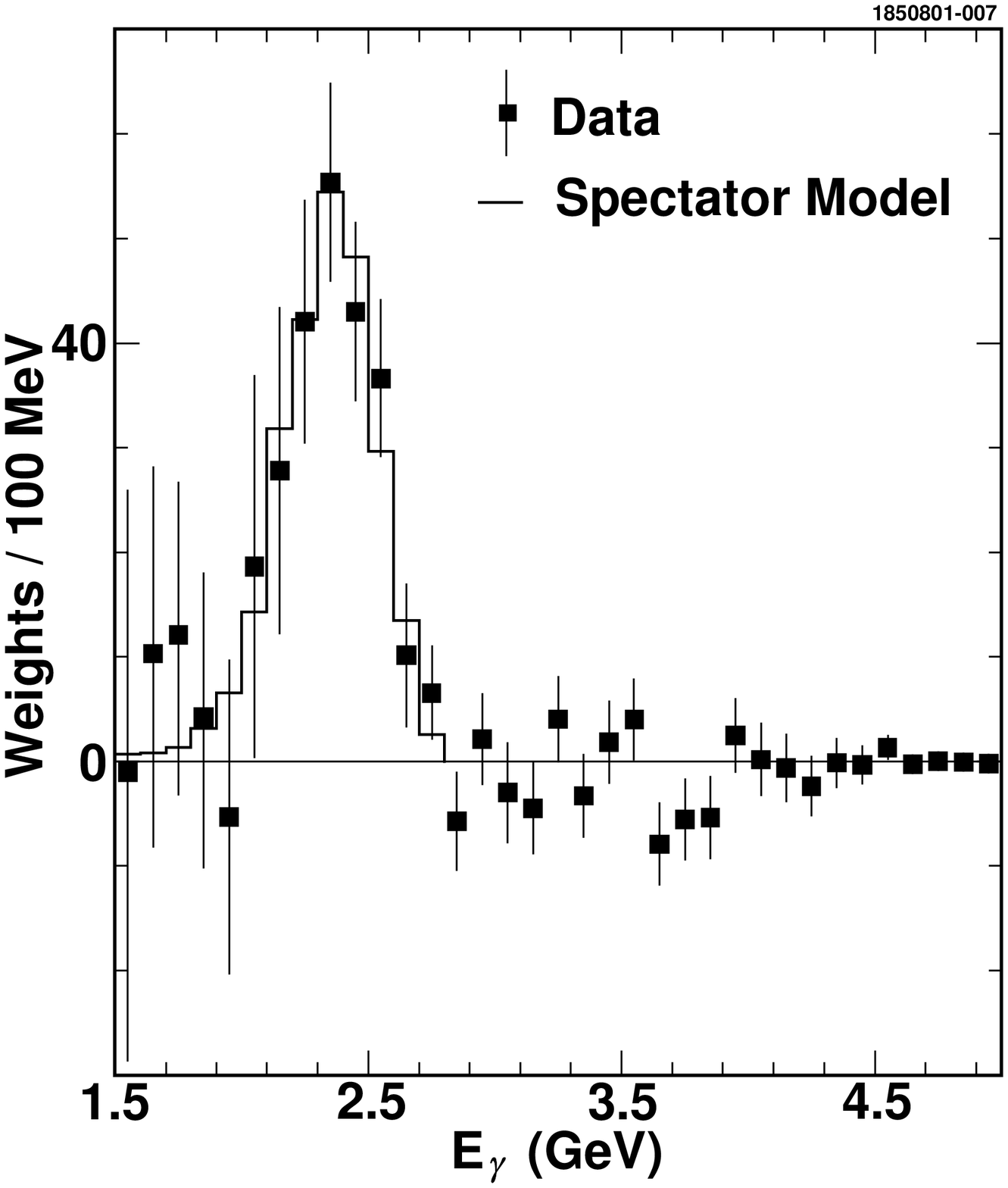}
   
    \includegraphics{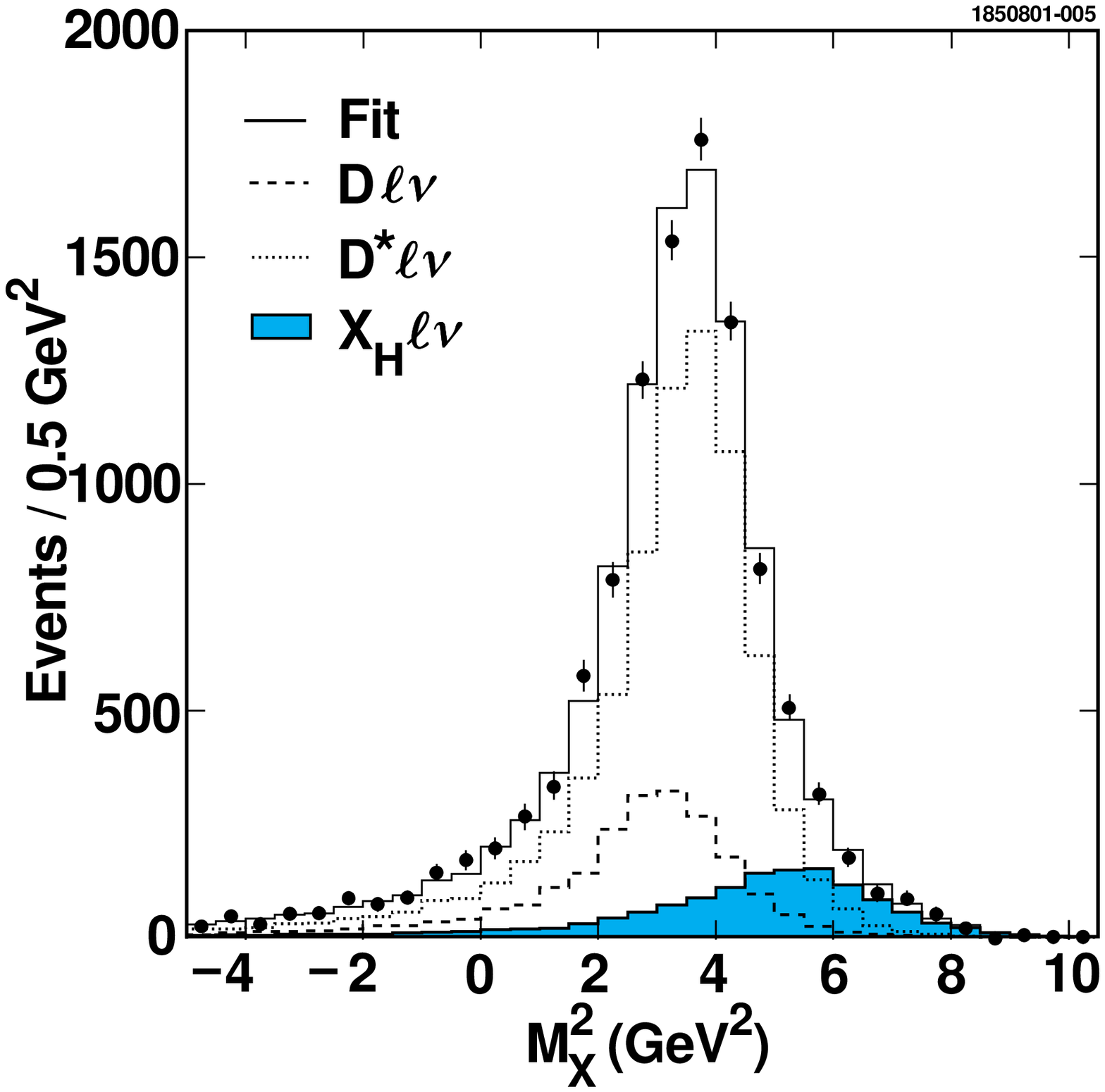}
    \caption{\it
      Distributions used as input to the preliminary CLEO ``moments''
      analysis for $V_{cb}$ from inclusive $b\to cl\nu$; on the
     left $E_\gamma$ is shown for CLEO
     $b\to s\gamma$ data along with a Monte Carlo prediction from
     a tuned spectator model, and on the right the hadronic mass
     for CLEO $B\to X_c\ l\nu$ events.
      \label{fig:egammx2}
      }
\end{figure}

\begin{figure}[htb]
  \vspace{6.9cm}
    \includegraphics{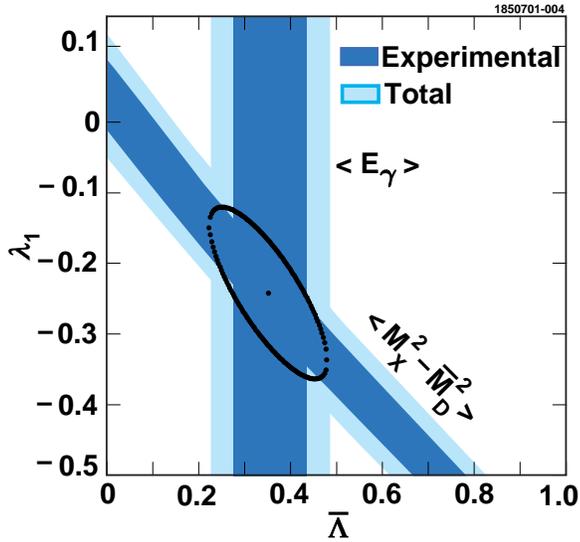}
   
    \caption{\it 68\% confidence level
      bands in the $(\bar{\Lambda},\lambda_1)$ plane from
      the preliminary CLEO measurements of the first moments
      of the distributions in Fig.~\ref{fig:egammx2}. The narrower
      bands include only experimental errors and the wider
      ones include $\sim 1/M_B^3$ theoretical uncertainty estimates.
      \label{fig:moments}
      }
\end{figure}

\begin{figure}[htb]
  \vspace{7.cm}
  \includegraphics{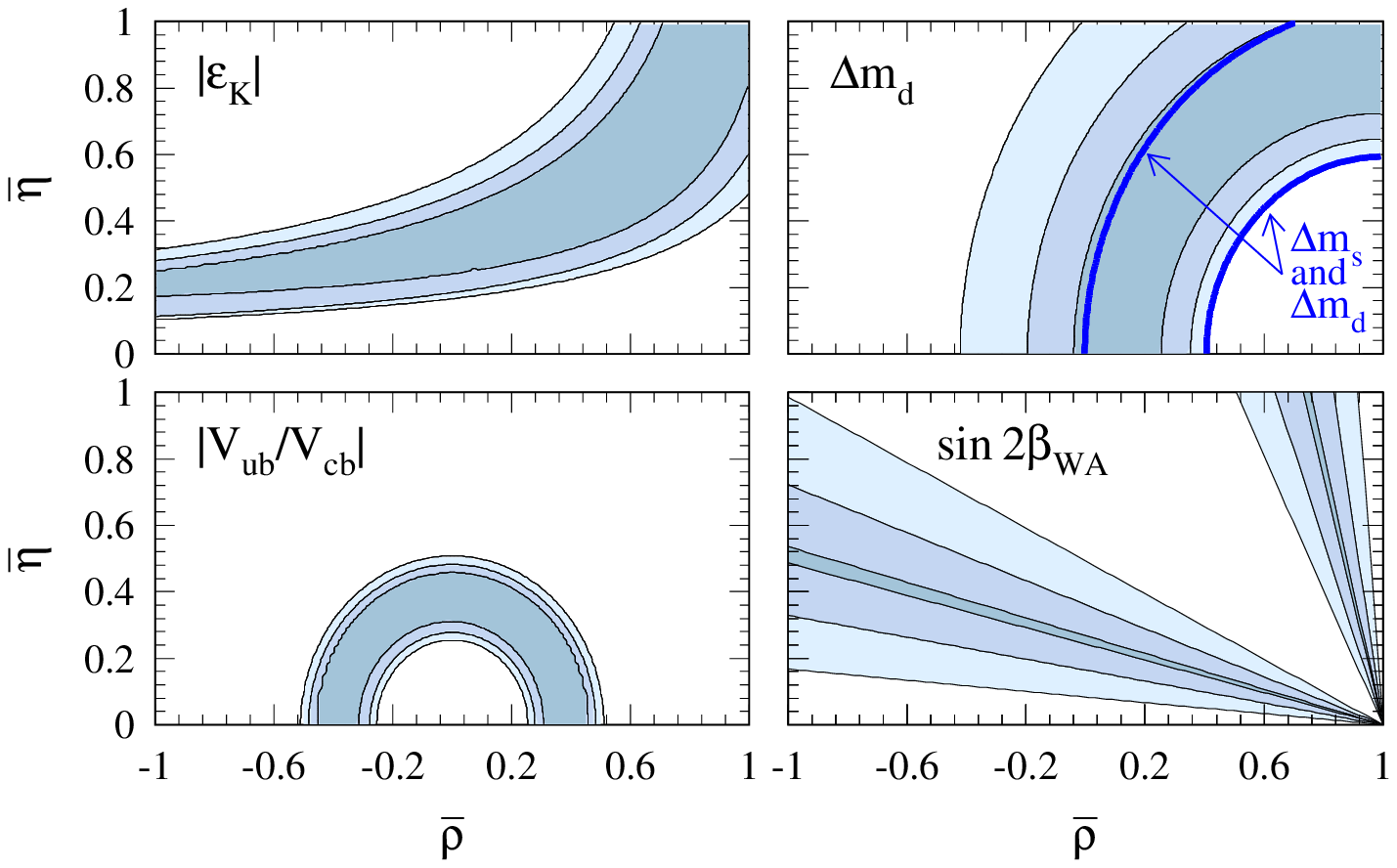}
  \caption{\it
    Frequentist CL's\cite{bib:HOCKER} 
     for different constraints in the $(\rhobar,\etabar)$
  plane. The bands contain regions 
  of $\ge$90\%, $\ge$32\%, and $\ge$5\%, respectively.
   [from ref.~8]
    \label{fig:ckmfour} }
\end{figure}

  CLEO has released an interesting new result\cite{bib:CLEOMX}
for $|V_{cb}|$ based on the combination of an
inclusive $b\to cl\nu$ analysis and the distributions in
two seemingly unconnected variables, the $\gamma$ energy
in $b\to s\gamma$ and the hadronic mass-squared 
in $B\to Xl\nu$.
The connection is made by HQET, which predicts\cite{bib:FLS}
the form of the function $h(\bar{\Lambda}, \lambda_1)$ in
$|V_{cb}|^2=\Gamma(b\to cl\nu)\times h(\bar{\Lambda}, \lambda_1)$,
in which $\bar{\Lambda}$ represents the mass of light quark and gluon
degrees of freedom and $-\lambda_1$ the average momentum-squared 
of the $b$-quark inside the $B$-meson. These
parameters, in turn, can be determined from the first
moments  $\left<E_\gamma\right>$ in $b\to s\gamma$ 
(sensitive to $\bar{\Lambda}$ only) 
and $\left<M_X^2-\overline{M}_{D}^2\right>$
in $B\to Xl\nu$, 
where $\overline{M}_{D}$ is the spin-averaged
$D$-mass, $\overline{M}_{D}=0.25M_D+0.75M_{D^*}$
(sensitive to both $\bar{\Lambda}$ and $\lambda_1$). The relevant
experimental distributions are shown in Fig.~\ref{fig:egammx2}.
The relation of these values to the HQET parameters is shown
in Fig.~\ref{fig:moments}; the values at the intersection of
the bands are then inserted into $h(\bar{\Lambda}, \lambda_1)$.
The result is $|V_{cb}|=(40.4\pm0.9\pm 0.5\pm0.8)\times 10^{-3}$,
with the errors covering uncertainties due to experimental
determination of $\Gamma$, experimental determination of 
$(\bar{\Lambda}, \lambda_1)$, and the theoretical accuracy for
$h$ ($1/M_B^3$
terms and scale $\alpha_s$), in that order. This result is subject to 
usual quark-hadron duality caveats associated with inclusive 
analyses.\cite{bib:ISGURBIG}

\section{Imposing Unitarity with Global Fits}

  Fig.~\ref{fig:ckmfour} shows the allowed bands in the 
$(\rhobar,\etabar)$-plane from four of the measured quantities.
Their overlap or lack thereof is a measure of consistency with the
Standard Model, and they can be combined to extract ``best-fit''
results for all the CKM-related quantities. The experimental 
and theoretical inputs themselves can be floated in the fits as well. 
Global fitting efforts which impose unitarity via constrained fits of 
current experimental and theoretical results start with all these
goals. But very different approaches and considerations of inputs 
result in quite different conclusions.

\begin{figure}[phtb]
  \vspace{6.8cm}
  \includegraphics{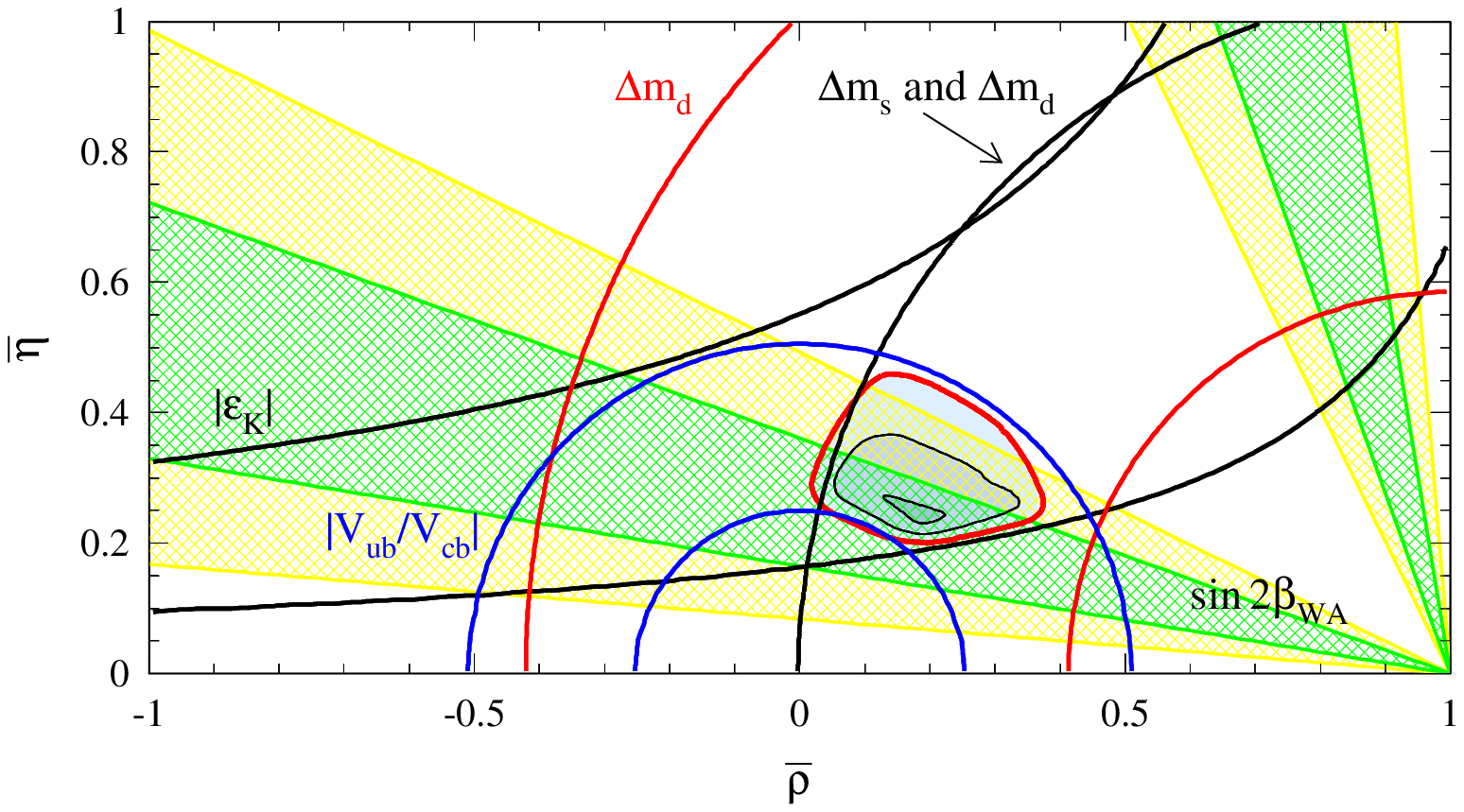}
  \caption{\it
    Frequentist CL's\cite{bib:HOCKER} for in the $\rhobar,\etabar$
  plane for the global fit of Hocker, \etal\cite{bib:HOCKER}\  
  The contours contain regions of $\ge$90\%, $\ge$32\%, and $\ge$5\%,
  respectively; $\ge$5\% levels are shown for individual constraints.
   [from ref.~8]
    \label{fig:hockerfit} }
\end{figure}

\begin{figure}[phtb]
  \vspace{6.8cm}
  \includegraphics{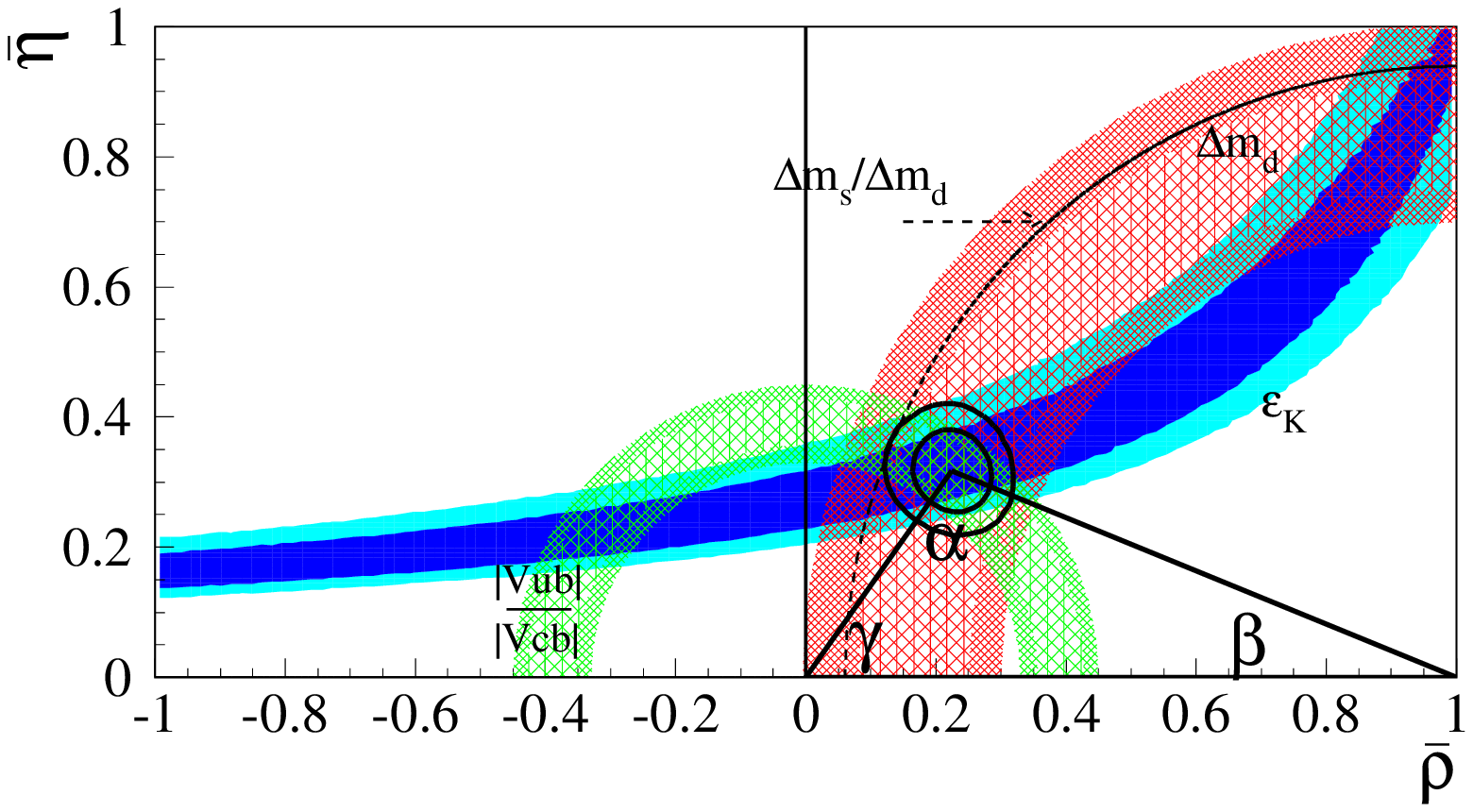}
  \caption{\it
    Confidence levels for in the $\rhobar,\etabar$
  plane for the global fit of Ciuchini, \etal\cite{bib:CIUCHINI} \ 
  Regions of $\ge$68\% and 95\% confidence level are shown by the
  contours (for the fit) and shaded areas (for the individual
  constraints).
   [from ref.~28]
    \label{fig:ciufit} }
\end{figure}

  Partial results of two of the more extensive global fitting efforts,
those of Hocker \etal\cite{bib:HOCKER} and Ciuchini \etal,\cite{bib:CIUCHINI}
are shown graphically in Figs.~\ref{fig:hockerfit} and \ref{fig:ciufit}.
Both fits use information available after the summer 2000 conference
season. While the final fit contours both center in the same general
$(\rhobar,\etabar)$ region, two significant differences between their 
analyses stand out:
a Bayesian vs. frequentist statistical approach to the problem, and the 
treatment of theoretical errors. The 
{\sl aggressive Bayesians} contend\cite{bib:CIUCHINI,bib:STOCCHI} that LQCD 
theoretical calculations are sufficiently mature to ascribe 
probabilistic meanings to central values and uncertainties no different
from experimental systematics; moreover they maintain that a Bayesian
procedure is entirely justified and results in a clearer interpretation 
of fit results. 
The counter-argument\cite{bib:HOCKER,bib:STONE} emphasizes that many of the
theoretical inputs are not well-known and therefore cannot be treated 
as having statistical or Gaussian errors. Furthermore, the 
{\sl conservative frequentist} viewpoint continues,
the more difficult theoretical calculations should be assigned an
{\sl allowed range, without any preferred central value}. For such
quantities the theoretical values {\sl should not contribute}
to the $\chi^2$, ranging freely over the entire allowed range
for any fixed point in fitting-parameter space. 
Also pointed out is a pernicious and 
subtle effect of Bayesian combination of as few as three variables 
specified to have
no preferred central value within a range: the Bayesian treatment can 
easily give a fallaciously narrowed and therefore misleading combined
fit. The latter effect occurs even using flat (rather than 
Gaussian) probability density functions.
The importance of these differences in approach will increase 
as more accurate data become available; it is conceivable 
that in addition to contrasting ``conservative'' vs. ``aggressive'' 
errors, they could come to conflicting conclusions on consistency with 
unitarity. These are not small considerations, and deserve further attention.

\section{The Role of CLEO-c}
\begin{figure}[htb]
  \vspace{4.1cm}
    \includegraphics{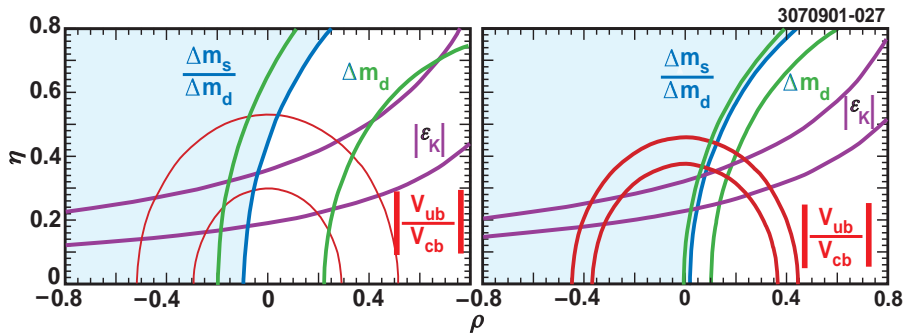}
    \caption{\it
      90\% confidence level contours in the $(\rho,\eta)$-plane
      from four measured observables, including current 
      theoretical and experimental errors (left) and
      after shrinking the relative theoretical uncertainties to the 2\%-level.
      \label{fig:imagine}
      }
\end{figure}

   Progress in resolving CKM metrology and unitarity consistency is
limited by a dearth of robust, accurate, and reliable 
non-perturbative QCD predictions. This point is made 
graphically in Fig.~\ref{fig:imagine}, which shows the dramatic effect 
on contours in the $(\rho.\eta)$ plane of arbitrarily shrinking 
contributions from theoretical errors to the 2\%-level. New
results from the $B$-factories will only serve to highlight the problem
by increasing the disparity between experimental and theoretical precision. 
LQCD is poised to provide the requisite relief in the coming years 
(Section~\ref{sec:QCD}). However, in order to have any
credibility, more accurate predictions must be confronted with equally 
accurate measurements. With only a handful of fundamental adjustable 
parameters (quark masses and the strong coupling constant), LQCD must 
be challenged to account for as large a set of observations within 
its domain as possible. At percent-level precision, there is 
now a paucity of the necessary experimental cross-checks available.

\begin{figure}[htb]
  \begin{center}
    \epsfig{file=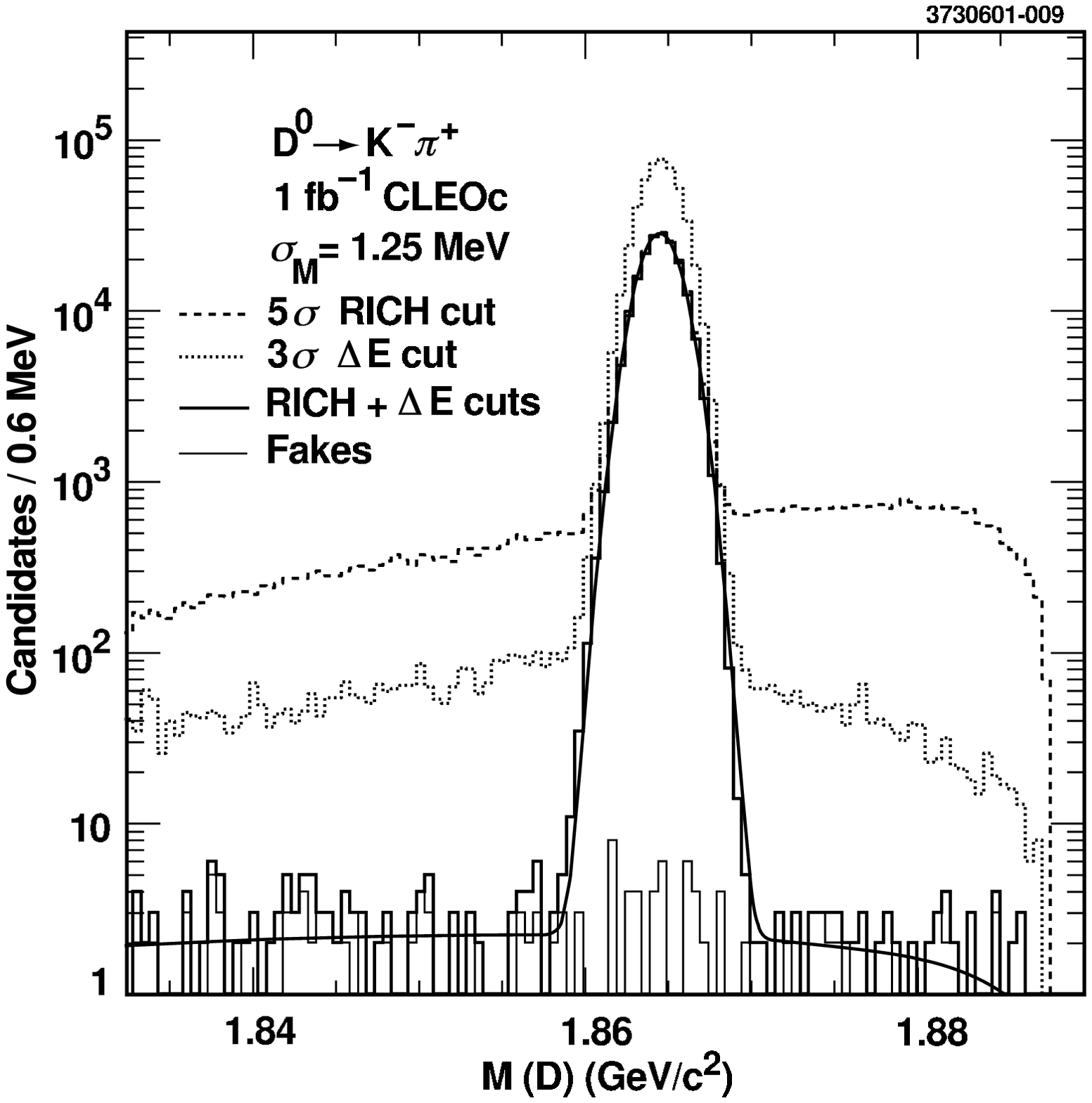,width=5.25cm}
    \hspace{0.5cm}
    \epsfig{file=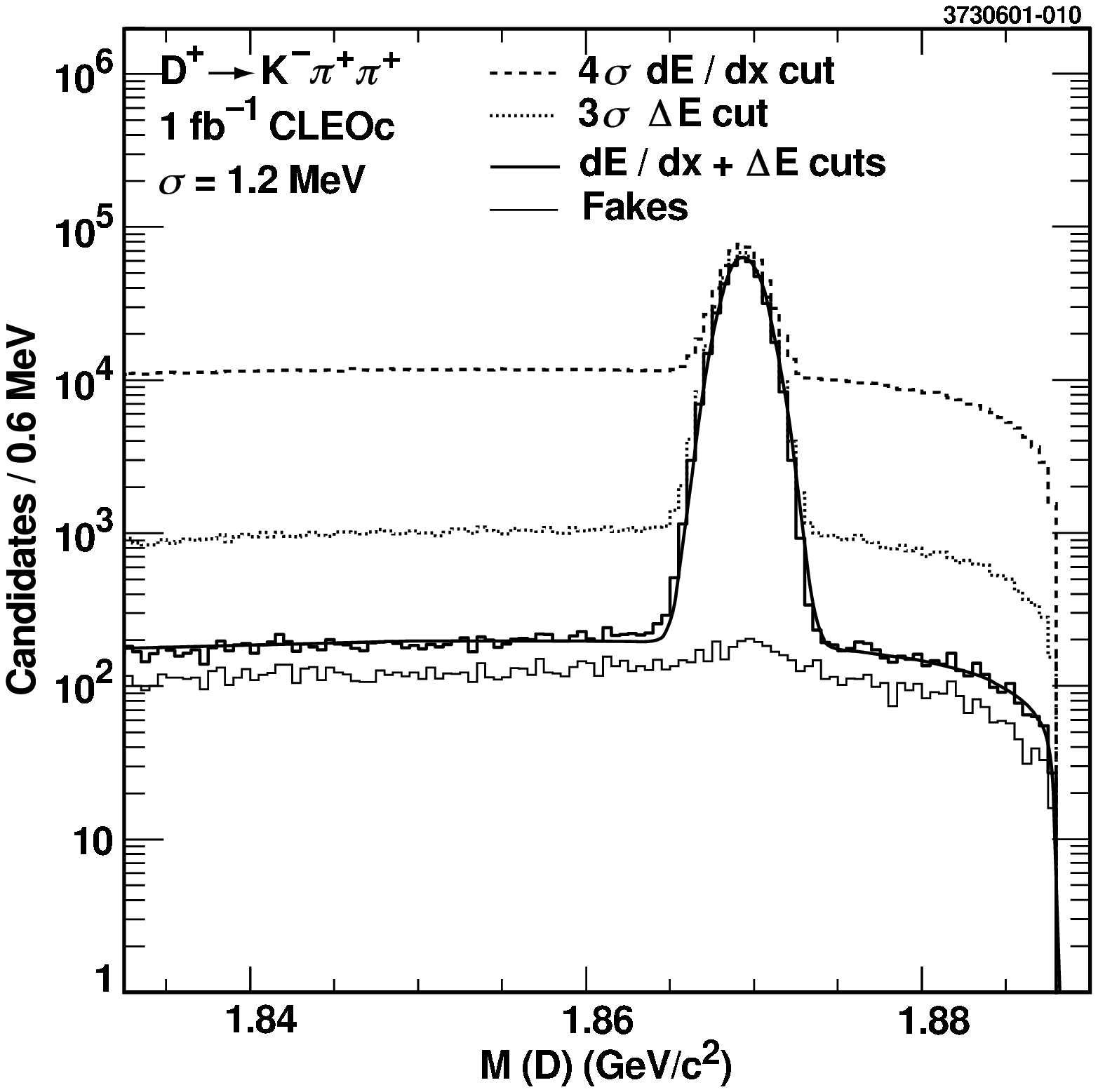,width=5.25cm}
    \caption{\it
      Reconstructed $K^-\pi^+$ and $K^-\pi^+\pi^+$
      $D$-candidate mass distributions for CLEO-c
      Monte Carlo events generated at $D\bar{D}$ threshold.
      \label{fig:d1tag}
      }
 \end{center}
\end{figure}

\begin{figure}[htb]
  \begin{center}
    \epsfig{file=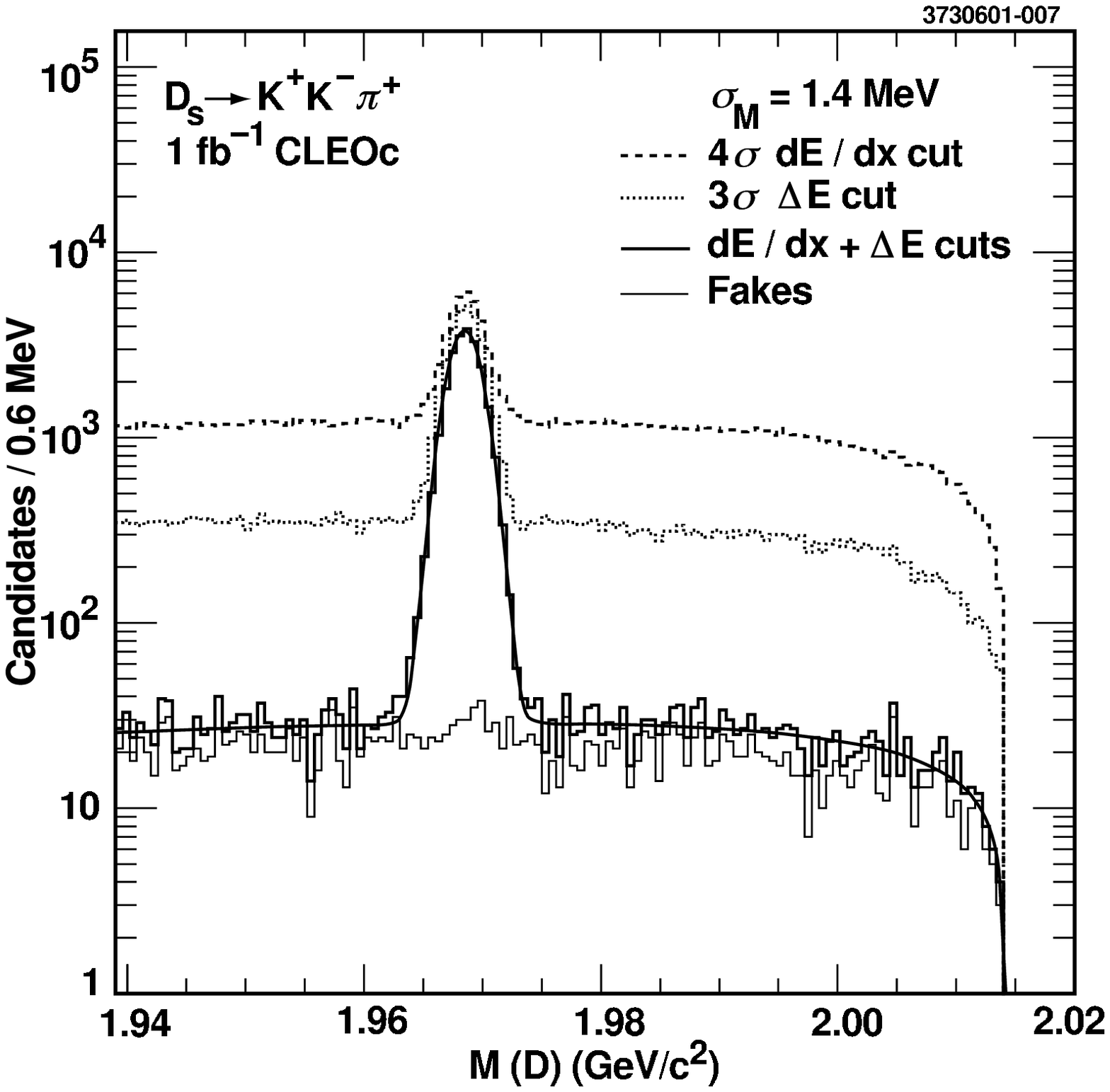,width=5.5cm}
    \hspace{0.5cm}
    \epsfig{file=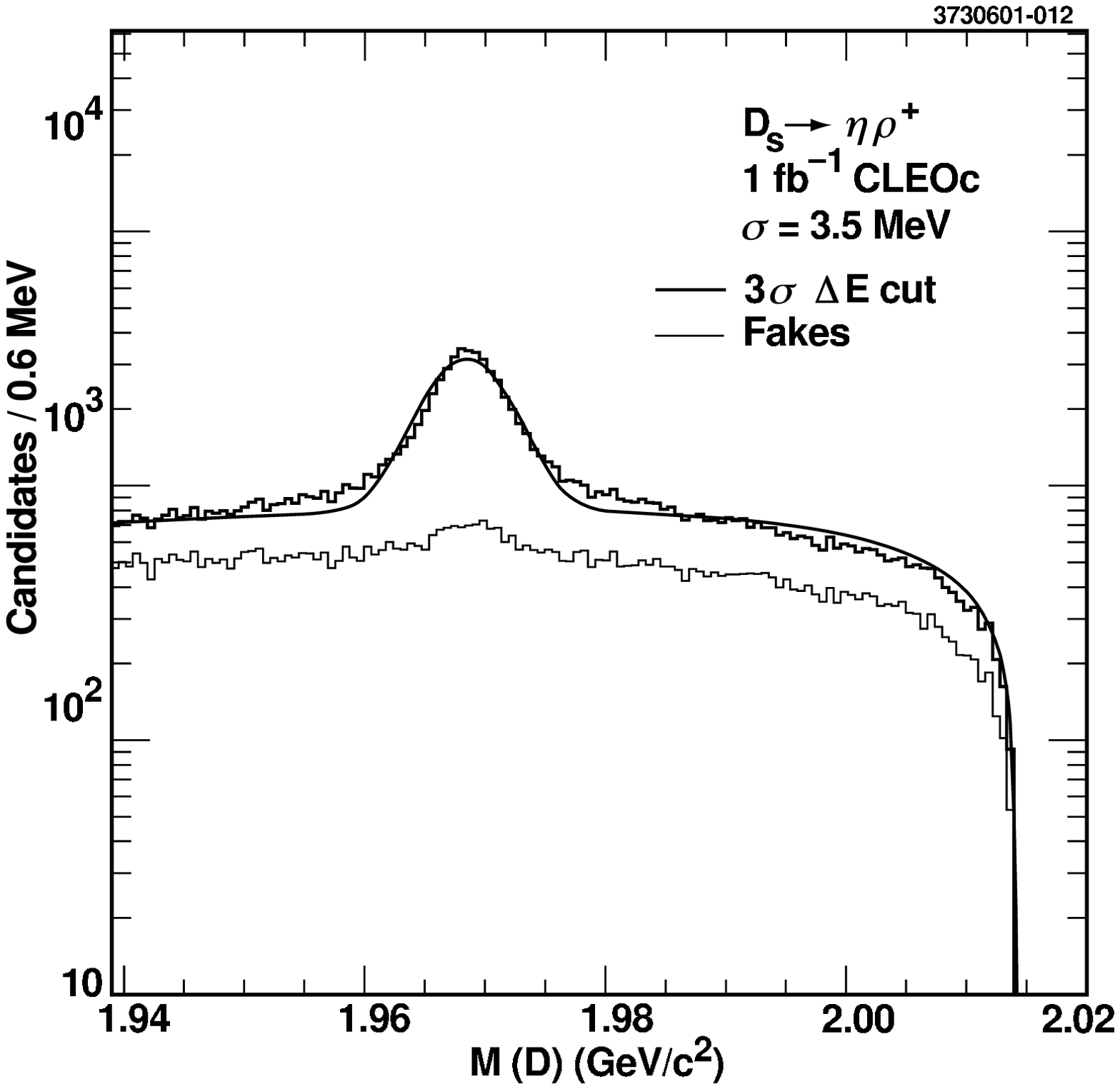,width=5.5cm}
    \caption{\it
      Reconstructed $K^+K^-\pi^+$ and $\eta\rho^+$
      $D_s$-candidate mass distributions for CLEO-c
      Monte Carlo events generated at $D_s\bar{D_s}$ threshold.
      \label{fig:ds1tag}
      }
 \end{center}
\end{figure}

\begin{figure}[phtb]
  \begin{center}
    \epsfig{file=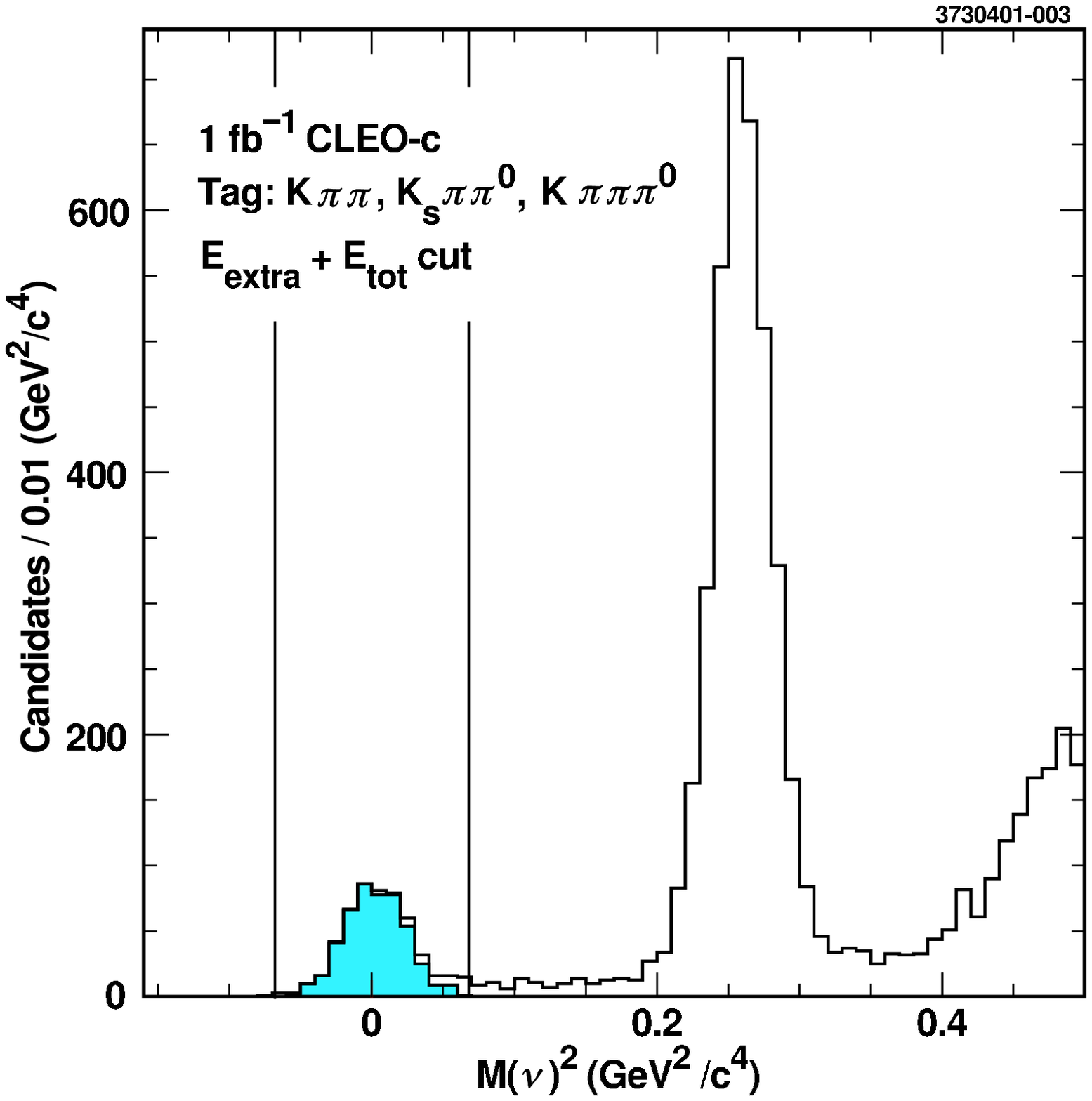,width=5.55cm}
    \hspace{0.5cm}
    \epsfig{file=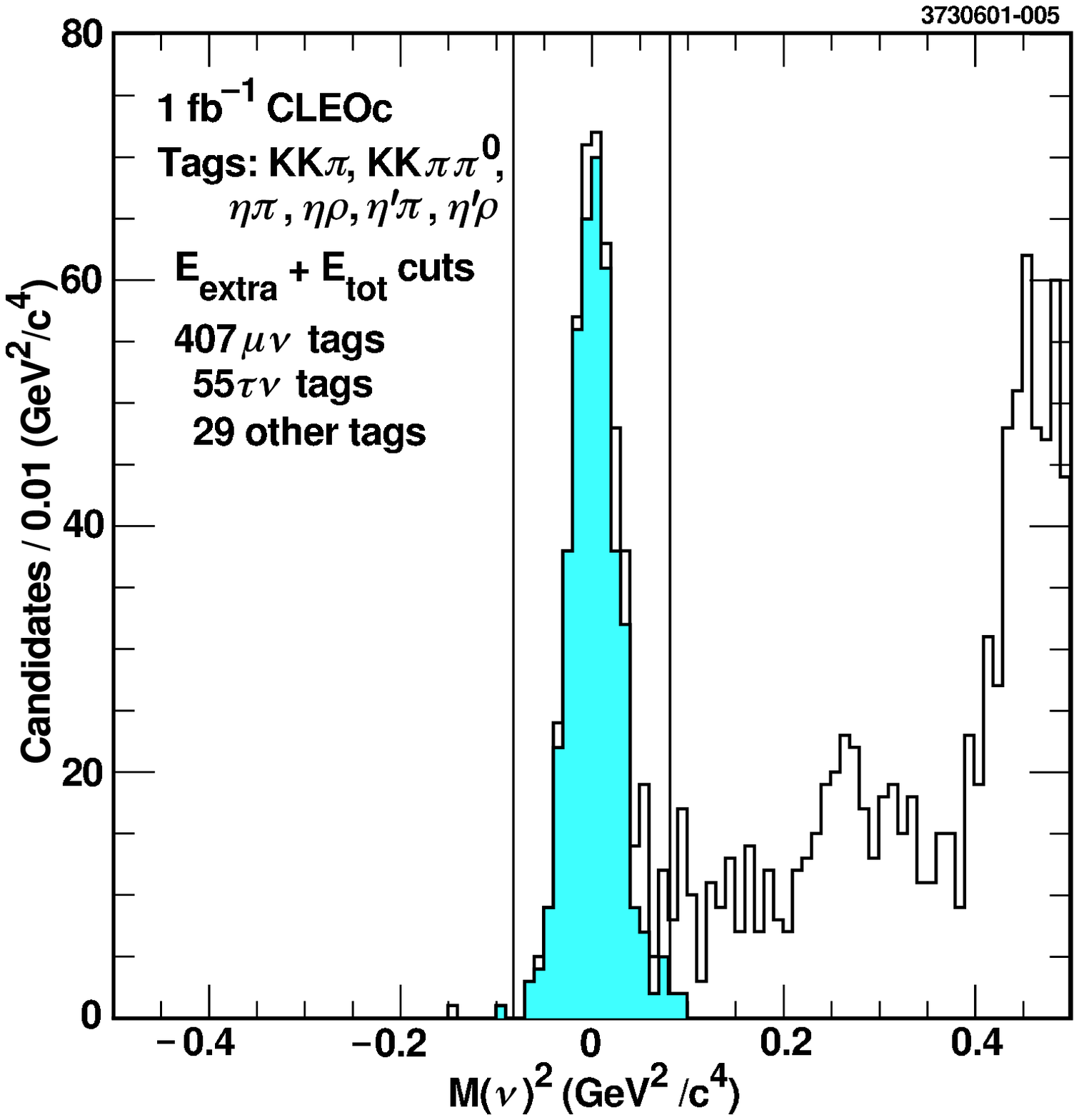,width=5.55cm}
    \caption{\it
      The missing-mass-squared distribution in
      tagged $D\to\mu\nu$ (left) and $D_s\to\mu\nu$ (right) events from
      CLEO-c Monte Carlo generated at $D\bar{D}$ and $D_s\bar{D_s}$ 
      thresholds, respectively. The shaded areas show correctly tagged
      candidates. The unshaded peak in the $D$ plot corresponds to
      $D\to K^0_L\mu\nu$ background.
      \label{fig:ddslep}
      }
 \end{center}
\end{figure}

\begin{figure}[phtb]
  \begin{center}
    \epsfig{file=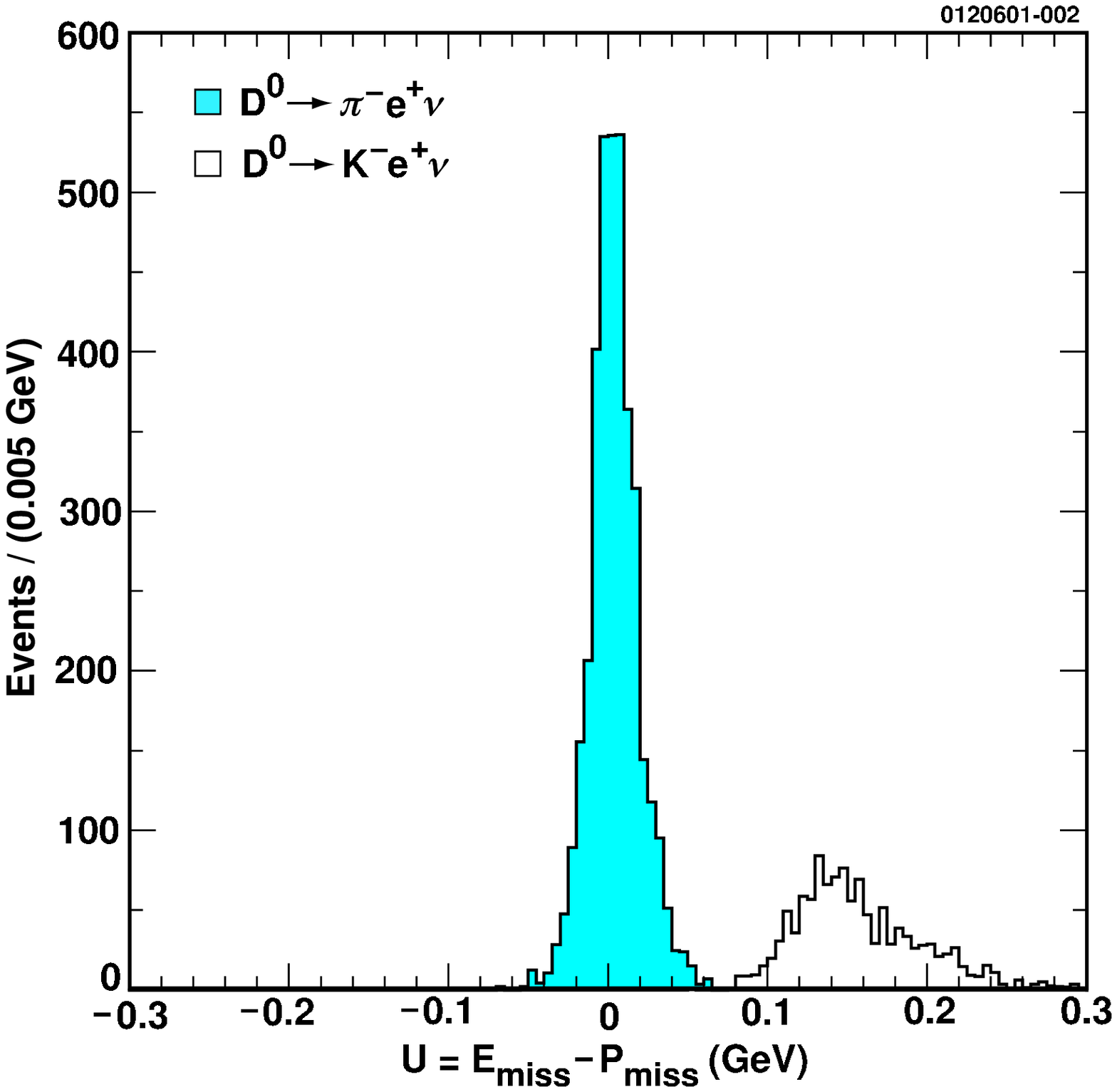,width=5.5cm}
    \hspace{0.7cm}
    \epsfig{file=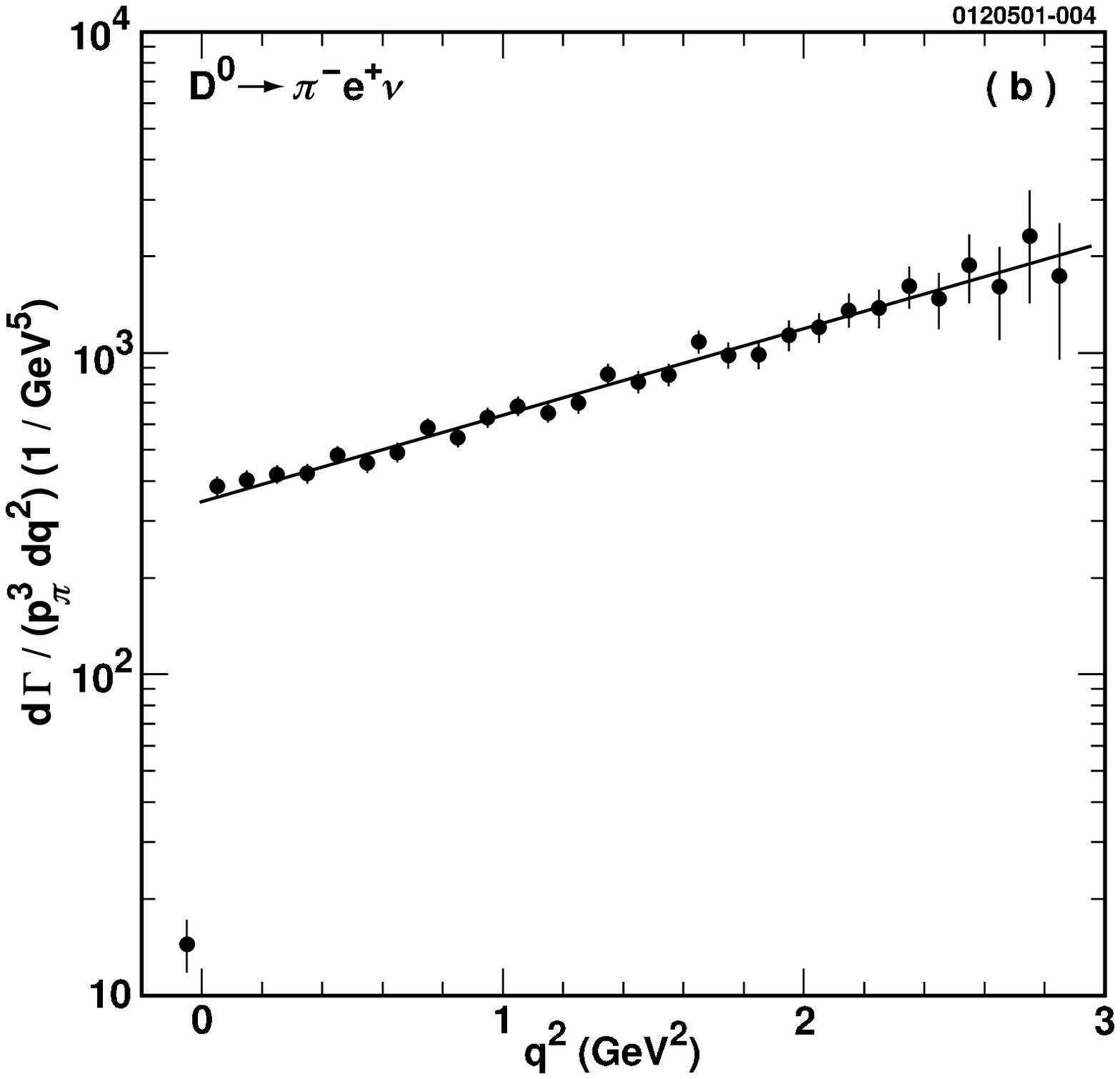,width=5.5cm}
    \caption{\it
      From CLEO-c Monte Carlo generated at $D\bar{D}$ threshold,
      distributions for tagged $D^0\to \pi^-e^+\nu$ candidates,
      in a missing-mass variable, shaded region
      being signal and unshaded misidentified $D^0\to K^-e^+\nu$
      background; and in $q^2$ with the overlayed
      fit for the form-factor. 
      \label{fig:dpilnu}
      }
 \end{center}
\end{figure}

\begin{figure}[phtb]
  \begin{center}
    \epsfig{file=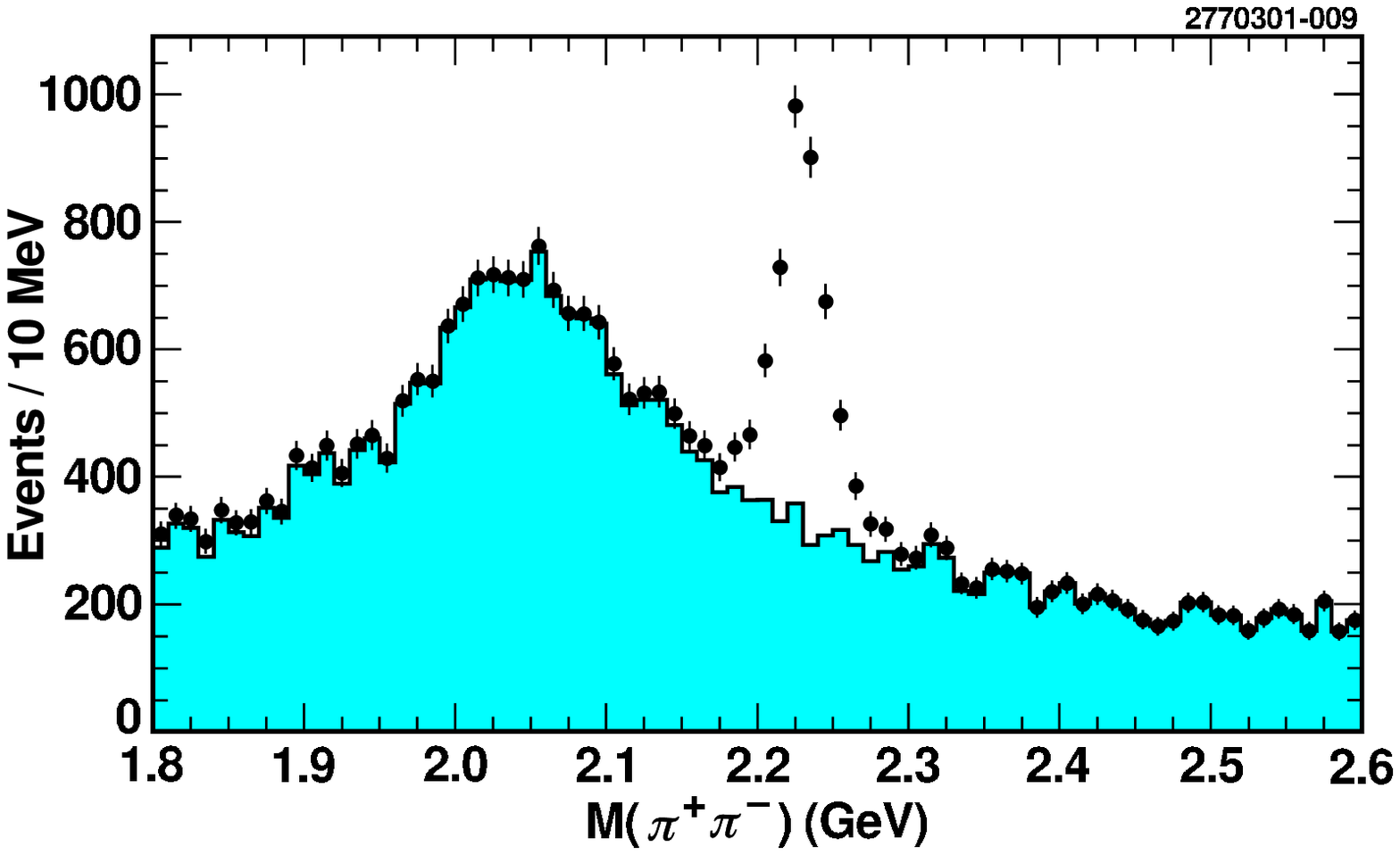,width=5.5cm}
    \hspace{0.7cm}
    \epsfig{file=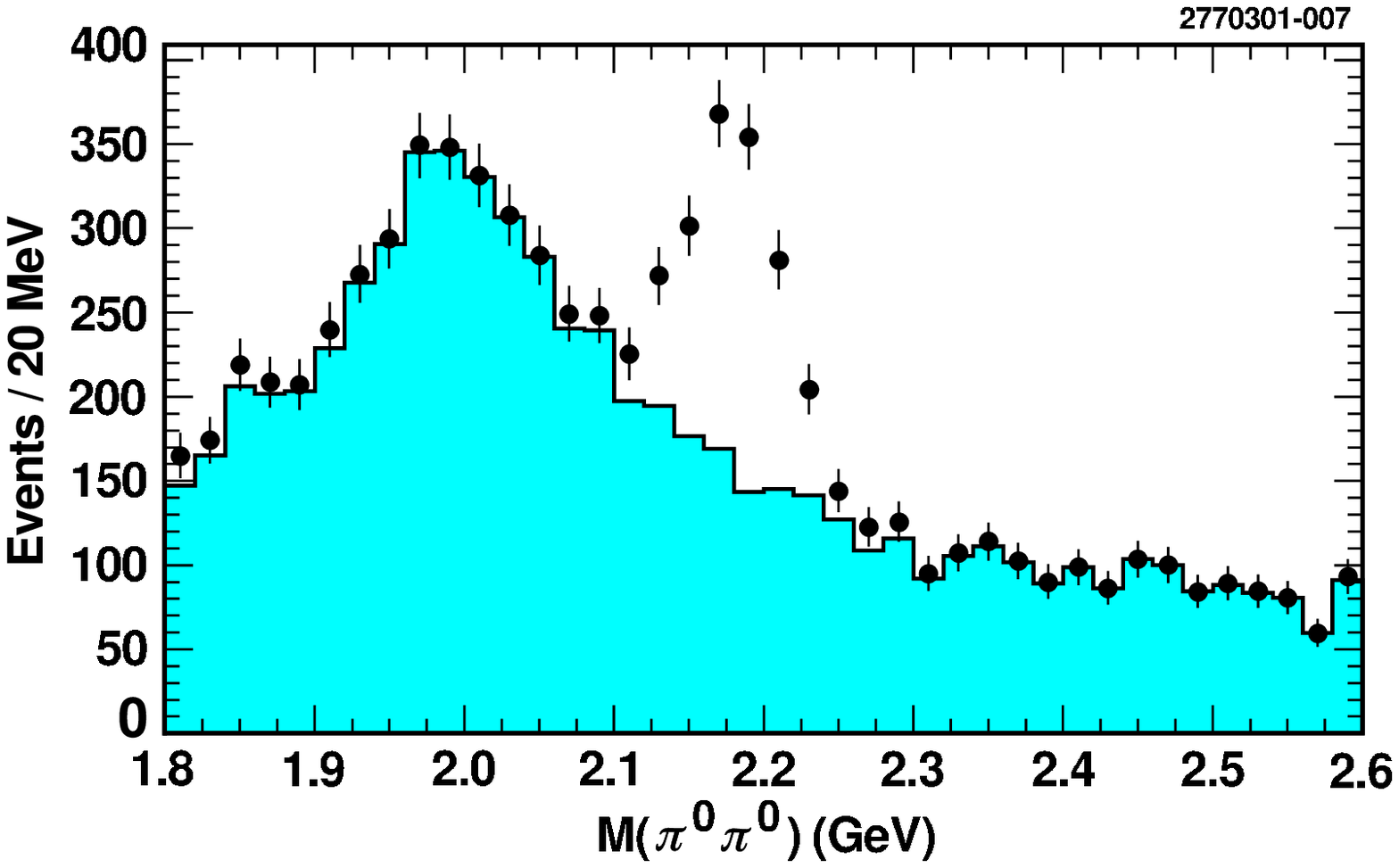,width=5.5cm}
    \caption{\it
      Two-particle mass distributions for $J/\psi\to\gamma\pi\pi$ candidates
      in CLEO-c Monte Carlo generated with $f_J(2220)$ 
      at the BES branching ratios 
      ${\cal B}(J/\psi\to\gamma f_J)\times{\cal B}(f_J\to\pi\pi)\sim
       5\times 10^{-5}$. Each plot corresponds to a sample of 
      $\sim$150M $J/\psi$ decays and has
      ${\cal B}(J/\psi\to\gamma f_4(2050))=2.7\times 10^{-3}$,
       ${\cal B}(f_4(2050)\to\pi\pi)=0.17$.
      \label{fig:g2pi}
      }
 \end{center}
\end{figure}

\begin{figure}[phtb]
  \begin{center}
    \epsfig{file=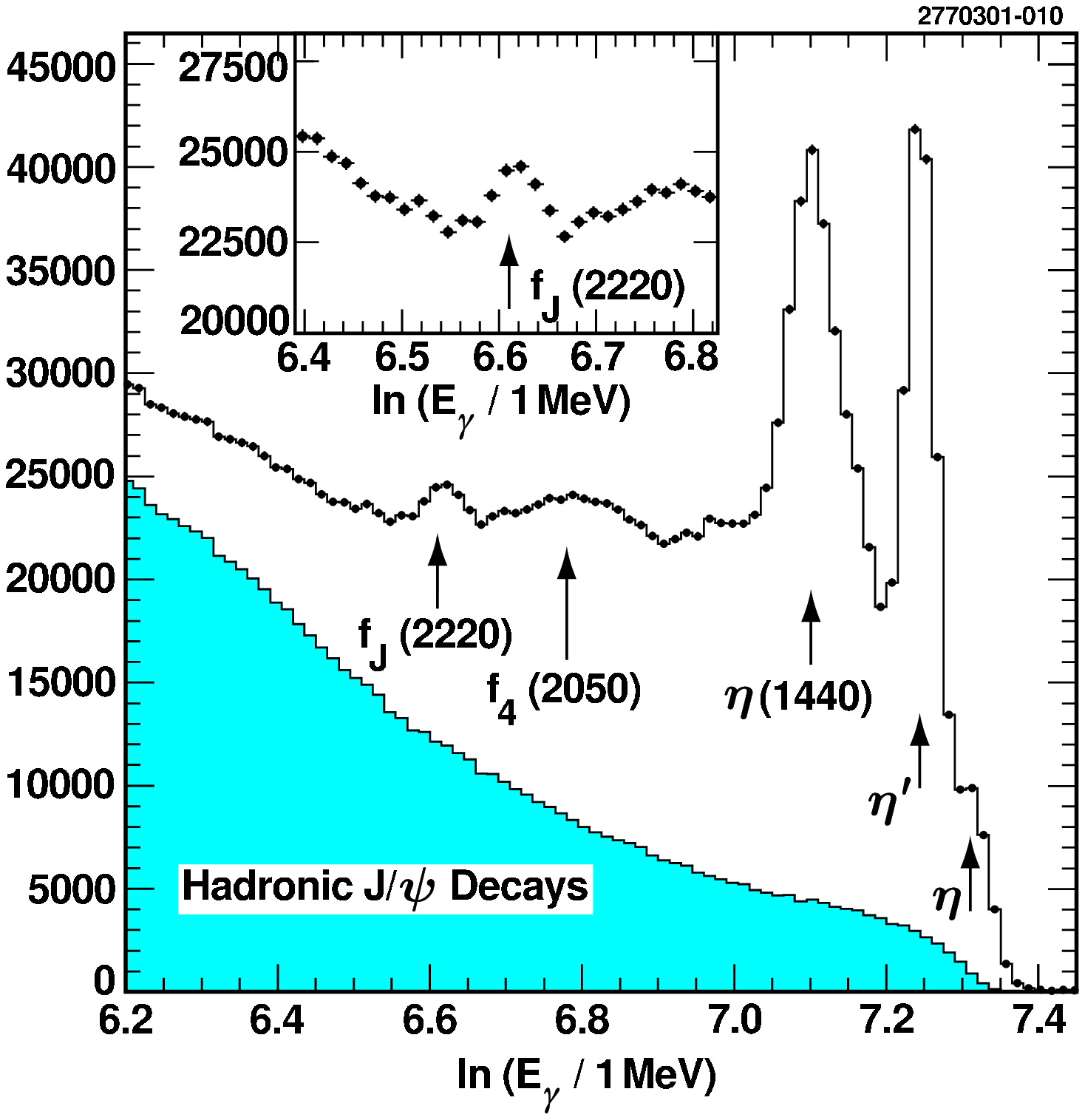,width=7cm}
    \caption{\it
      Inclusive photon spectrum in CLEO-c Monte Carlo of
      60M $J/\psi$ decays, showing sensitivity a $f_J(2220)$ signal
      with ${\cal B}(J/\psi\to \gamma f_J(2220))=8\times 10^{-4}$ 
      in relation to backgrounds with
      ${\cal B}(J/\psi\to \gamma f_4(2050)=2.7\times 10^{-3}$
      and ${\cal B}(J/\psi\to \gamma \eta(1440))=4.5\times 10^{-3}$.
      \label{fig:ginc}
      }
 \end{center}
\end{figure}

  The CESR-c/CLEO-c\cite{bib:CLEOC} program is designed to fill this 
experimental void on an ideal time-scale. The plan calls
for $e^+e^-$ datasets of unprecedented size in the 
tau-charm and $\Upsilon$ energy regimes with a modern, well-seasoned 
detector over a five year period. Only modest enhancements to the 
CESR collider, in the form of new superconducting wiggler magnets, are 
necessary to run in the center-of-mass energy range from $\sqrt{s}$=3-11~GeV
with peak luminosity scaling as $\sim s$,
${\cal L}\approx 0.15-1.2\times 10^{33}$~cm$^{-2}$~sec$^{-1}$. 
Few changes to the CLEO~III detector, infrastructure, 
and analysis tools are required; CLEO-c will easily be the most capable 
detector to have run in this energy range. Datasets of 10$^9$ events at 
the $J/\psi$, 3~fb$^{-1}$ at each of $D\bar{D}$ and $D_s\bar{D_s}$ 
thresholds, and 1~fb$^{-1}$ at each of $\Upsilon(1S)$, $\Upsilon(2S)$, 
and $\Upsilon(3S)$ resonances will provide precise benchmark
measurements with which to confront emerging LQCD calculations.

  The decay constants for $D$ and $D_s$ can be measured accurately
at pair-production thresholds by flavor-tagging with a fully-reconstructed 
known decay and looking for the $\mu\nu$ (and $\tau\nu$ for $D_s$) of 
the appropriate charge and kinematics, thereby measuring
the branching fraction. The cleanliness of single-tag $D$ and $D_s$ 
samples is shown in Fig.~\ref{fig:d1tag} and \ref{fig:ds1tag}. 
Fig.~\ref{fig:ddslep}
shows missing-mass-squared distributions for tagged $D,D_s\to\mu\nu$ 
samples from 1~fb$^{-1}$ at each threshold; signal events are copious
and well-separated from dangerous backgrounds, projecting to 
$\le$4\% uncertainty these branching fractions. Combined with
the corresponding lifetime errors of $\le$2\% and $|V_{cj}|$ errors of 
$\le$1\% obtained from unitarity, relative uncertainties of $\sim$2\% are
expected for both $f_D$ and $f_{D_s}$.

  Another strength of CLEO-c is its ability to isolate and measure
semi-leptonic decays cleanly, as shown for $D\to\pi l\nu$ in
Fig.~\ref{fig:dpilnu}. Such measurements allow the determination
of form-factor $shapes$ for comparison with theory, and $|V_{cj}|$
using the measured branching fractions along with normalization from theory.
These shapes and normalizations can be cross-checked with
a large variety of exclusive semi-leptonic decay modes, with
uncertainties expected in the few percent level. The ratio 
$|V_{cd}|/|V_{cs}|$, a more straightforward test for 
theory, will have experimental uncertainties in the 1-5\% range.

  A second vexing issue for heavy flavor physics is the increasing importance
of having accurate $D$ branching fractions, which are beginning to
limit corresponding $B$-physics results. CLEO-c can address this
issue as well, by measuring the rate of double- to single-tags for
the desired modes. Once again, threshold measurements are seen to be
extremely clean and accurate. Branching fractions will be measured
to $\sim$1\% or better for $D^0\to K^-\pi^+$ and $D^+\to K^-\pi^+\pi^+$ 
for $D$'s and $\sim$2\% for $D_s\to\phi\pi$; 
correspondingly large improvements
over current uncertainties are expected for other modes as well.

  A second crucial part of the CLEO-c/CESR-c program is to spend time
studying the charmonium and bottomonium states, where several
QCD prizes await. Probing radiative $J/\psi$ decays for the
on-again, off-again $f_J(2220)$ should prove decisively whether it
exists\cite{bib:PDG} and, if so, its status as $q\bar{q}$, glueball, 
or hybrid. Evidence of the $f_J(2220)$ is 
conflicting,\cite{bib:BES1} but has 
been reported in $K\bar{K}$, $\pi^+\pi^-$, $\pi^0\pi^0$,
and $p\bar{p}$ final states. Fig.~\ref{fig:g2pi} shows CLEO-c's sensitivity
for two of these modes in exclusive $J/\psi\to\gamma f_J(2220)$ when
produced at the observed BES branching fractions in a 
small fraction of the projected CLEO-c dataset of 10$^9$ $J/\psi$'s. 
If it exists anywhere near this level, it will be seen and a partial-wave
analysis will measure its quantum numbers $J^{PC}$. The inclusive
photon spectrum, as shown in Fig.~\ref{fig:ginc}, will also prove
to be a powerful tool in discovery or confirmation of any narrow
resonance produced in radiative $J/\psi$ decays. If the $f_J(2220)$
is truly a glueball, it should not be seen in the $\sim$20~fb$^{-1}$
of CLEO~II and CLEO~III data (or Belle or BaBar's either) as a 
product of two-photon collisions; CLEO has already set a limit\cite{bib:CLEOFJ}
based on 4.8~fb$^{-1}$.

  Other interesting quarkonia physics topics which will provide more
much-needed grist for the LQCD mill, include discovery and
mass-measurement of the $\eta_b$ (ground state $^1S_0$ of $b\bar{b}$),
$h_b$ ($^1P_1$), glueballs in glue-rich radiative $\Upsilon$ decays,
glue-quark hybrid states, scalar resonances with or without
glue content in radiative $J/\psi$ decays, and many, many more.

  One might well ask how the above capabilities stack up against
that expected from BES, or against $B$-factories with
$\sim$400~fb$^{-1}$. While it is always risky to bet what someone else is 
not capable of, charm physics at the $\Upsilon (4S)$ faces some
obstacles that improve only slowly or not at all with more data. 
Prime among barriers is the absence of the kinematic constraints
available at threshold. However, extrapolating from 
previous CLEO efforts, keeping such obstacles and potential ways 
around them in mind, yields optimistically three
times larger errors on charm decay constants and branching fractions. If these
estimates prove to instead be pessimistic, it will still be true that 
$\Upsilon(4S)$ charm analyses will be systematics-limited in a way that is
complementary to CLEO-c. The other competition targeting tau-charm physics
is BES~II at the BEPC storage ring in Beijing. Currently BEPC operates
at more than an order of magnitude lower luminosity than that projected
for CESR-c, and the proposed machine and detector upgrades to
``factory'' levels could be completed by 2005 at the earliest.
CLEO-c plays a unique and crucial role in the years ahead.

\section{Outlook}

  The continuing productive interplay among accelerator development,
theoretical insight, and experimental technique has brought
CKM physics to the threshold of a major transition, one that
is qualitative as well as a quantitative. The $B$-factories 
are operating beyond expectations, promising a rich array of new
measurements. Foremost among those would be observation of $CP$ 
violation in $B$-decays, which can happen with data already in-hand, 
assuming Nature cooperates. The fortuitously-timed maturation of LQCD 
along with a reality-grounding CLEO-c experimental program should provide 
crucial breakthroughs in unraveling the effects of strong interactions. 
The approach to {\sl precision} CKM physics on both experimental
and theoretical fronts is underway.

\section{Acknowledgments}

  Our hosts for {\sl Physics in Collision 2001} and its
international and local organizing committees are to be
congratulated for organizing a productive and stimulating
conference at Seoul National University.
The patient assistance of my CLEO colleagues was
essential in the preparation of this presentation, and greatly appreciated.

%

%

\end{document}